\documentclass[twocolumn]{aastex6}

\usepackage{natbib}
\usepackage{amsmath}
\usepackage{multirow}
\usepackage{graphicx}
\usepackage{color}
\usepackage{threeparttable}
\usepackage{enumitem}
\usepackage{mathtools}
\usepackage{lipsum}
\def\maxij1820{MAXI~J1820+070}
\def\gx339{GX~339--4}
\def\exo1846{EXO~1846--031}
\def\at2019wey{AT2019wey}

\def\rxte{{RXTE}}
\def\xmm{{XMM-Newton}}

\def\swift{{Swift}}
\def\nustar{{NuSTAR}}
\def\nicer{{NICER}}

\def\reltrans{{\tt reltrans}}

\begin{document}
%\title{Reverberation lags in black hole low-mass X-ray binaries in the \textit{NICER} archive}
%\title{A systematic study of reverberation lags in black hole X-ray binaries with \textit{NICER}}
\title{The NICER ``Reverberation Machine": A systematic study of time lags in black hole X-ray binaries}
\author{Jingyi~Wang\altaffilmark{1},
Erin~Kara\altaffilmark{1},
Matteo~Lucchini\altaffilmark{1}, 
Adam~Ingram\altaffilmark{2,3}, 
Michiel van der Klis\altaffilmark{4}, 
Guglielmo~Mastroserio\altaffilmark{5}, 
Javier~A.~Garc{\'\i}a \altaffilmark{5,6}, 
Thomas~Dauser\altaffilmark{6}, 
Riley~Connors\altaffilmark{5}, 
Andrew~C.~Fabian\altaffilmark{7},
James~F.~Steiner\altaffilmark{8},
Ron~A.~Remillard\altaffilmark{1},
Edward~M.~Cackett\altaffilmark{9}, 
Phil~Uttley\altaffilmark{4}, 
Diego~Altamirano\altaffilmark{10}.
}
\affil{
\altaffilmark{1}MIT Kavli Institute for Astrophysics and Space
Research, MIT, 70 Vassar Street, Cambridge, MA 02139, USA\\
\altaffilmark{2}Department of Physics, Astrophysics, University of Oxford, Denys Wilkinson Building, Keble Road, Oxford OX1 3RH, UK\\
\altaffilmark{3}School of Mathematics, Statistics and Physics, Newcastle University, Herschel Building, Newcastle upon Tyne, NE1 7RU, UK\\
\altaffilmark{4}Astronomical Institute, Anton Pannekoek, University of Amsterdam, Science Park 904, NL-1098 XH Amsterdam, Netherlands\\
\altaffilmark{5}Cahill Center for Astronomy and Astrophysics, California Institute of Technology, Pasadena, CA 91125, USA\\ 
\altaffilmark{6}Remeis Observatory \& ECAP, Universit\"{a}t
Erlangen-N\"{u}rnberg, 96049 Bamberg, Germany\\ 
\altaffilmark{7}Institute of Astronomy, University of Cambridge, Madingley Road, Cambridge, CB3 0HA\\
\altaffilmark{8}Harvard-Smithsonian Center for Astrophysics, 60 Garden St., Cambridge, MA 02138, USA \\
\altaffilmark{9}Department of Physics \& Astronomy, Wayne State University, 666 W. Hancock St, Detroit, MI 48201, USA\\
\altaffilmark{10} School of Physics and Astronomy, University of Southampton, Highfield, Southampton, SO17 1BJ\\
}

\begin{abstract}
We perform the first systematic search of all \nicer\ archival observations of black hole (and candidate) low-mass X-ray binaries for signatures of reverberation. Reverberation lags result from the light travel time difference between the direct coronal emission and the reflected disk {\color{black}component}, and therefore their properties are a useful probe of the disk-corona geometry. We detect new signatures of reverberation lags in 8 sources, increasing the total sample from 3 to 11, and study the evolution of reverberation lag properties as the sources evolve in outbursts. We find that in all of the 9 sources with more than 1 reverberation lag detection, the reverberation lags become longer and dominate at lower Fourier frequencies during the hard-to-soft state transition. This result shows that the evolution in reverberation lags is a global property of the state transitions of black hole low-mass X-ray binaries, which is valuable in constraining models of such state transitions. The reverberation lag evolution suggests that the corona is the base of a jet which vertically expands and/or gets ejected during state transition. We also discover that in the hard state, the reverberation lags get shorter, just as the QPOs move to higher frequencies, but then in the state transition, while the QPOs continue to higher frequencies, the lags get longer. We discuss implications for the coronal geometry and physical models of QPOs in light of this new finding.
\end{abstract}
\keywords{accretion, accretion disks --- 
black hole physics}

\section{Introduction} \label{intro}
Black holes are the most extreme laboratories to study accretion and ejection physics, and to eventually test theories of gravity. In addition to the supermassive black holes at the centers of galaxies, there are also stellar-mass black holes, which are mostly discovered when they are X-ray bright in systems called X-ray binaries (XRBs). These are binary systems comprised of a compact object accreting material from a companion star (see \citealp{remillard2006x} for a review and references therein). If the mass of the companion star is lower than roughly one solar mass, it is called a low-mass XRB (LMXB; e.g., \citealp{1983ARA&A..21...13B}). Depending on the nature of the central compact object, the source is categorized as either a black hole XRB (BHXB) or a neutron star XRB (NSXB).

\begin{figure*}
\centering
\includegraphics[width=1.\linewidth]{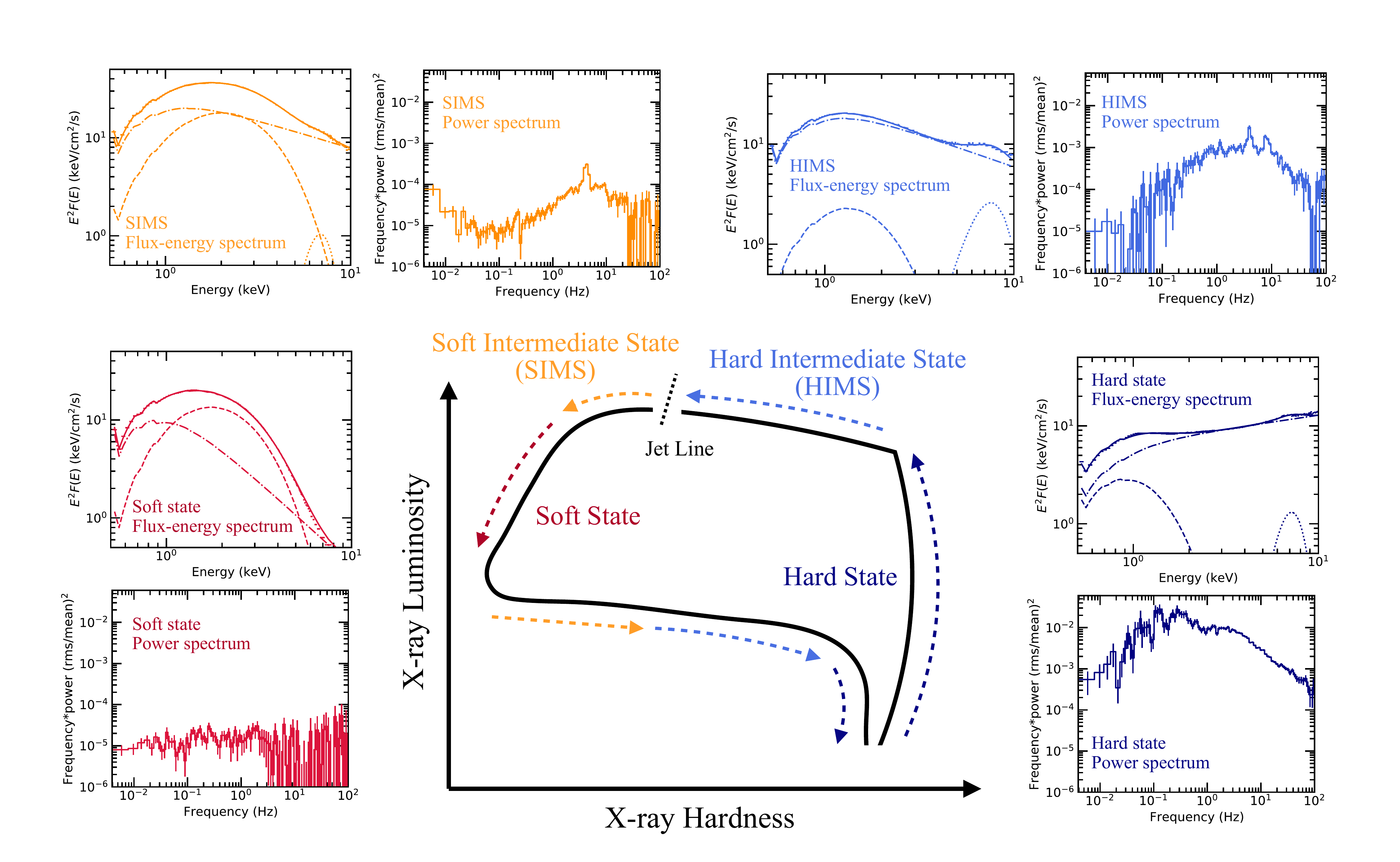}
\caption{{\color{black}The hardness-intensity diagram (HID), where the arrows show the typical temporal evolution of BHLMXBs through the 4 different states (hard state, HIMS, SIMS, and soft state) represented by different colors. The HID is encompassed by representative flux-energy spectrum and power spectrum of each state, using the \nicer\ data of the BHLMXB \maxij1820. The flux-energy spectrum (solid line) is fitted phenomenologically with the sum of a thermal disk emission (\texttt{diskbb}, dashed line), the coronal emission approximated by a power-law (dash-dotted line), and the relativistic iron line approximated by a Gaussian line (dotted line). }
}
\label{fig:hid_schematics}
\end{figure*}

Typically, LMXBs are transient sources which go through cycles of activity called outbursts (\citealp{2005Ap&SS.300..107H,remillard2006x,tetarenko2016watchdog}; and see {\color{black}Fig.~\ref{fig:hid_schematics}}). During the full outburst of a given black hole low-mass X-ray binary (BHLMXB), the system starts from the quiescent state with a very low mass accretion rate, and increases in luminosity through the ``hard state", when X-ray emission is dominated by a hard X-ray component that can be approximately described by a power-law with photon index $\Gamma$ {\color{black}in the range of 1.4--1.8} and is usually attributed to the non-thermal coronal emission. After several weeks, the spectrum can change dramatically, and as the BHXB transitions, it progresses through the ``hard intermediate state" (HIMS), to the ``soft intermediate state" (SIMS), before switching to the ``soft state" dominated by the thermal emission of the accretion disk {\color{black}(see Fig.~\ref{fig:hid_schematics}}). The fractional rms can reach 30\%--40\% in the hard state, but drops to only a few percent in the soft state. After spending {\color{black}weeks to months} in the soft state, the luminosity steadily drops, and the source makes a transition back to the hard state (at a lower luminosity than the hard-to-soft transition) and then eventually decays to quiescence. {\color{black}There is a remarkable coupling between accretion and ejection in BHLMXBs, as the properties of the radio jets strongly depend on the accretion states \citep{fender2004towards,fender2009jets}. In the hard state, a compact and mildly relativistic jet is almost always observed (e.g. \citealt{dhawan2000scale}); and there is a correlation between the radio and X-ray luminosities \citep{hannikainen1998most,corbel2000coupling,gallo2003universal}.} %Notably, there is a correlation between the radio and X-ray luminosities in the hard state
%, suggesting that the {\color{black}inflowing and outflowing} material are coupled to each other \citep{hannikainen1998most,corbel2000coupling,gallo2003universal}. 
During the hard-to-soft transition, the radio luminosity decreases, suggesting that the compact jet is slowly quenched, and at the so-called ``jet line" (which loosely corresponds to the boundary between the HIMS and SIMS), a highly relativistic, discrete and ballistic jet is launched \citep{bright2020extremely}. {\color{black}During the soft-to-hard transition and the decay in the hard state, when the compact jet is re-formed is not well understood (e.g., \citealp{miller2012disc,kalemci2013complete}).}

Though we have observed the ubiquitous non-thermal X-ray emission from XRBs for several decades, our understanding of its nature is still limited \citep{done2007modelling,nowak2011corona}. The widely accepted idea is that the X-ray corona, a hot plasma located close to the black hole with electron temperature of the order of $\sim100$~keV and optical depth $\tau\approx 1$, produces the power-law-like X-rays via thermal Comptonisation of the disk photons. However, how the X-ray corona is generated in the first place, and how it evolves throughout an outburst, are not well understood. The most {\color{black}frequently used} coronal models include: (1) an inner hot flow (e.g., advection-dominated accretion flow; \citealp{esin1997advection,narayan2008advection}) interior to a truncated thin accretion disk \citep{mahmoud2019reverberation,zdziarski2021accretion}, and (2) the corona being the base of a jet \citep{markoff2001jet,markoff2005going} or an aborted jet \citep{ghisellini2004aborted}, the simplest realization of which is the lamppost geometry \citep{matt1991iron,matt1992iron,dauser2013irradiation}.

It then becomes advantageous to study the spectral-timing evolution of XRBs to be able to probe the inner disk-corona region. With a technique called reflection spectroscopy, we model the time-averaged flux-energy spectrum of the disk material that is irradiated by the coronal photons. The most prominent reflection features include the Fe K emission complex ($6-7$~keV), the Fe K-edge ($\sim 7-10$\,keV), and the Compton hump ($\sim 20-30$\,keV), all broadened by relativistic effects \citep{fabian1989x,matt1993iron,garcia2014_relxill}. However, in modeling the flux-energy spectrum alone, {\color{black}limitations arise} from {\color{black}degeneracies such as the model choice for the coronal emission \citep{dzielak2019comparison}, and the positive correlation between the iron abundance and the BH spin parameter \citep{javier_gx339}. These prevent} us from unequivocally answering important questions, such as how truncated is the disk in the bright hard state (e.g., \citealp{wang2020evolution}) or how quickly is the BH spinning \citep{dauser2013irradiation}.

X-ray timing analyses can help break some of these degeneracies and build a more complete picture by providing insights about the nature of variability. Fourier timing techniques allow us to separate variability on different timescales that are believed to result from distinct physical processes \citep{uttley2014x}. Besides its unique signature in the flux-energy spectrum, the reflection process can also be seen in the form of a time lag feature, which is commonly referred to as reverberation. It arises because of the light travel time delay between the direct coronal emission and the reflected emission. Reverberation lags have been detected in both AGN and BHXBs at relatively high Fourier frequencies, in the forms of variations in the soft-excess emission and/or the Fe K emission line lagging behind those in the energy band dominated by the coronal emission (e.g., \citealp{fabian2009broad, kara2016global, de2017evolution}). At relatively low Fourier frequencies (when probing longer timescale variability), there is a ubiquitous hard lag whose amplitude is too large to result from light travel time difference; the most popular model is mass accretion rate fluctuations propagating through {\color{black}the disk, moving from the colder to the hotter regions}, causing soft photons to respond before hard {\color{black}photons} \citep{lyubarskii1997flicker}. 

In a typical lag-frequency spectrum, the low-frequency hard lag {\color{black}(positive by convention)} has a power-law-like shape {\color{black}and decreases with frequency}, making it possible to see the soft lags {\color{black}(negative by convention) caused by reverberation} at high frequencies. The soft lag flips into positive above some frequency $\sim 1/(2\tau_0)$ due to phase-wrapping, where $\tau_0$ is the intrinsic time delay between the coronal and reflected emissions. At even higher frequencies, the lag oscillates {\color{black}around zero}, again as a result of phase-wrapping (see e.g., Fig.~7 in \citealp{emmanoulopoulos2014general}, and also Fig.~\ref{fig:gx339}e). The observed lag amplitude depends both on $\tau_0$ and the level of ``dilution". {\color{black}The lag-frequency spectrum is measured between two energy bands, each \textit{dominated} by the coronal or reflected emission. However, in reality, both energy bands {\color{black}contain contributions from coronal and reflected emission}. This dilutes the time delay causing the observed lag amplitude to be smaller than the intrinsic lag (see e.g., \citealp{kara2013closest,uttley2014x,de2015tracing}).}
%-- this is the dilution effect. This causes the observed lag amplitude to be smaller than the intrinsic lag (see e.g., \citealp{kara2013closest,uttley2014x,de2015tracing}).} %Dilution is caused by the flux in the two energy bands between which we measure the lags not being contributed exclusively by either coronal or reflected emission, respectively (see e.g., \citealp{kara2013closest,uttley2014x,de2015tracing}).
%Dilution is caused by the energy bands between which we measure the lags not being entirely dominated by coronal or reflected emission

In addition to the time lags, another observable that may encode information about the inner disk-corona geometry {\color{black}is} quasi periodic oscillations \citep[{\color{black}QPOs}; see the review][and references therein]{ingram2019review}. Type-C QPOs are {\color{black}strong and narrow peaks in the power spectrum. They are prominent in the bright hard state and HIMS, with centroid frequency rising from a few mHz to $\sim10$~Hz as the sources evolve. Type-B QPOs are characterized by a relatively strong and narrow peak with centroid frequencies of 5--6~Hz. The most rare Type-A QPOs are weak and broad in the power spectrum. Type-B and Type-A QPOs appear in the SIMS \citep{2005Ap&SS.300..107H}.} Therefore, a common definition of the SIMS is the presence of the Type-B and/or Type-A QPOs. One promising interpretation of the Type-C QPOs invokes the relativistic precession model, in which the fundamental frequency is simply assumed to be the Lense-Thirring precession frequency at some characteristic radius \citep[e.g., the disk truncation radius within which the inner hot flow is precessing;][]{ingram2009low}. As the Type-B QPOs are observed to be closely related to the launch of discrete jet ejections \citep{homan2020rapid}, they are likely to have a jet-based origin, e.g., the jet is launched from a precessing flow and precesses with it \citep{motta2015geometrical,stevens2016phase,de2019systematic}.

A systematic study of reverberation lags has been performed for both Seyfert galaxies \citep{de2013discovery, kara2016global} and BHXBs \citep{de2015tracing}. Intrinsically, as the number of counts within the variability timescale over which we detect reverberation lags is much lower in BHXBs than in Seyfert galaxies, reverberation lag detection is more challenging for BHXBs. With the \xmm\ archival data of 10 BHXBs, soft reverberation lags were detected in two sources in the hard state: \gx339\ and H1743--322 \citep{de2015tracing}. 

Now with the Neutron Star Interior Composition Interior Explorer \citep[\nicer;][]{gendreau2016neutron} we have the opportunity to {\color{black}systematically search for reverberation signatures in BHXBs, and to globally study their evolution.}
%take BHXB reverberation studies to the next level. 
NICER is an ideal instrument for spectral-timing analysis because of its large effective area in the 0.3--12~keV range, superior timing capabilities (i.e.{\color{black},} high resolution and very little dead-time), with negligible pileup and low background. In addition, \nicer\ has fast slew capabilities, which allows us to monitor BHXB systems at high cadence, throughout their outbursts. NICER has observed several outbursts, some from newly discovered black holes, covering all spectral states. Therefore, with the rich \nicer\ archive, we can perform a systematic search for reverberation lags. 

Thus far, reverberation lags have been reported in two BHXBs with NICER. In the hard state, reverberation lags are detected in \maxij1820\ \citep{kara2019corona} and {\color{black}in} \gx339\ \citep[{\color{black}during} its 2017 and 2019 outbursts;][]{wang2020relativistic}. In \citet{wang2021disk}, we were able to trace reverberation lags into the HIMS for the first time, thanks to the unprecedented data quality of \maxij1820. Our main finding was that the Fourier frequency range of the reverberation signature decreases during the hard-to-soft state transition with an increasing lag amplitude, possibly due to a vertically expanding corona. We successfully tested this hypothesis with the publicly-available reverberation model \reltrans\ \citep{ingram2019public,mastroserio2021modelling}, whereby the longer time lags during the state transition are explained with an increasing height of the lamppost corona over the accretion disk. This X-ray coronal expansion preceded a radio flare by $\sim 5$~days, hinting that the corona is the base of the jet, and the hard-to-soft transition is marked by the corona expanding vertically and launching a jet knot that propagates along the jet stream at relativistic velocities. These X-ray reverberation results were confirmed independently by \citet{de2021inner}.

Motivated by these reverberation results on individual sources, in this paper we perform the first systematic search for reverberation lags in BH and BH candidate (BHC) LMXBs observed with \nicer. We perform a {\color{black}detailed} study on their evolution throughout the hard and intermediate states, to learn whether the behavior seen in \maxij1820 is common to other sources. The lag properties we focus on are which Fourier frequency range is dominated by reverberation and what the reverberation lag amplitudes are. This paper is organized as follows. Section~2 describes the observations (sample and data reduction) and our method to efficiently search for reverberation lags. Section~3 focuses on the search results and the evolution of reverberation lags. We present the discussion in Section~4, and summarize the results in Section~5.

\section{Observations and method} \label{obs}

\subsection{The sample}\label{sample}
The systems of interest in our study are the BH and BHC LMXBs. We cross-matched the sample in \citealt{tetarenko2016watchdog} (see Tables 8 and 12 therein) and included newly discovered systems since then (private communications with John Tomsick). In total, since its launch in 2017 until June 2021, \nicer\ has observed 26 BH/BHC LMXBs (see Table~\ref{tab:sample_gg} and \ref{tab:sample_no_gg} for the full list).

\begin{table*}[htb!]
\begin{center}
\caption{The BH/BHC LMXB sample that is observed by \nicer\ and has good data groups resulting from our reverberation machine, sorted by the peak count rate. \label{tab:sample_gg}}
\footnotesize
\begin{tabular}{cccccccc}\hline \hline
Name& Peak count rate$^1$ & Peak rms$^1$ &Exposure$^1$ &States  & $\#$ of good & $\#$ of soft & Grouping\\
& (10$^3$~counts/s) & ($\%$) & (ks) & achieved$^2$  & groups  & lag detection & criteria$^3$ \\
\hline
MAXI J1820+070 & 61 & 24 & 419.8 & \textbf{H},\textbf{I},S & 88 & 84 & [0.3, 0.3, 0.5, 250]\\% & 3115 
MAXI J1348--630 & 29 & 21 &132.9 & \textbf{H},\textbf{I},S  & 34 & 15& [0.3, 0.3, 0.5, 250]\\%& 1283 
MAXI J1535--571 & 17 & 14 &73.9 & \textbf{H},\textbf{I},S  & 14 & 12 & [0.3, 0.3, 0.5, 250]\\%& 1048 
MAXI J1631--479 & 4.7 & 8 &80.2 & \textbf{I},S& 12 & 9& [0.3, 0.3, 0.5, 250]\\%& 932 
GX 339--4 & 4.5 & 27 &172.5 & \textbf{H},\textbf{I},S  & 14 & 9& [0.3, 0.3, 0.5, 250]\\% & 746
MAXI J1803--298 & 4.0  & 10 & 42.6 &  \textbf{H},I,S & 2 & 1& [0.3, 1, 1, 200] \\% & 236
MAXI J1727--203 & 3.3 & 12 &21.5 & H,\textbf{I},S & 3 & 2& [1, 1, 1, 150]\\%  & 200
EXO 1846--031 & 0.99 & 16 &85.5 & \textbf{H},\textbf{I},S & 3 & 2& [1, 1, 1, 100]\\% & 181 
AT2019wey & 0.78 & 18 &382.0 & H,\textbf{I} & 22 & 3& [0.3, 0.5, 0.5, 250] \\%& 355
GRS 1915+105 & 0.32 & 18 & 133.3 & H,\textbf{I}  & 3 & 2 & [0.3, 0.3, 0.5, 250]\\
\hline
\end{tabular}
\\
\raggedright{\textbf{Notes.} \\The count rates are for $0.3-12$~keV, and are normalized for 52~FPMs. The rms is calculated also in $0.3-12$~keV, with the Fourier frequency range of 1--10~Hz. $^1$These values are calculated among the data in the good groups. As we do not have good groups in the soft state (for their low variability) where count rate is usually higher than the hard state, and also with great \nicer\ coverage, the peak count rate and exposure could be underestimated compared to the values taking into account all the archival data. The peak rms could also be underestimated when there is no good group in the faint hard state (for their low count rate). $^2$The hard state, intermediate state, and the soft state are represented by H, I, and S respectively; the bold indicates where the soft lag is detected. $^3$The first three values correspond to the fractions within which the relative changes of the hardness ratio, the count rate, and the fractional variance are in each group (by default, 0.3, 0.3, and 0.5 respectively); the last value is the minimal ${\rm SNR}_{\Delta t, ind}$ we count as a good group. See text in Section~\ref{method} for more details.}
\end{center}
\end{table*}

\subsection{Data reduction} \label{reduction}

The \nicer\ data are processed with the data-analysis software NICERDAS version v2020-04-23\_V007a, and energy scale (gain) release ``optmv10". We use the following filtering criteria: the pointing offset is less than $60\arcsec$, the pointing direction is more than $30^\circ$ away from the bright Earth limb, and more than $15^\circ$ away from the dark Earth limb, and the spacecraft is outside the South Atlantic Anomaly (SAA).  Data are required to be collected at either a sun-angle $>60^\circ$ or else collected in shadow (as indicated by the ``sunshine’’ flag).  We filter out commonly-noisy detectors FPMs \#14, 34, and 54. In addition, we flag any ``hot detectors" in which X-ray or undershoot {\color{black}(detector resets triggered by accumulated charge)} rates are far out of line with the others ($\sim 10 \sigma$) and exclude those detectors for the GTI in question.  We select events that are not flagged as ``overshoot" {\color{black}(typically caused by charged particle passing through the detector and depositing energy)} or ``undershoot" resets (EVENT\_FLAGS=bxxxx00), or forced triggers (EVENT\_FLAGS=bx1x000), and require an event trigger on the slow chain {\color{black}which is optimized for measuring the energy of the event (i.e., excluding fast-chain-only events where the fast chain is optimized for more precise timing).} A ``trumpet" filter on the ``PI-ratio’’ is also applied to remove particle events from the detector periphery \citep{bogdanov2019constraining}. The resulting cleaned events are barycenter corrected using the \texttt{FTOOL} \texttt{barycorr}. The background spectrum is estimated using the 3C50 background model \citep{remillard2021empirical}. GTIs with overshoot rate $>2$~FPM$^{-1}$~s$^{-1}$ are excluded to avoid unreliable background estimation. We use the RMF version ``rmf6s" and ARF version ``consim135p", which are both a part of the CALDB xti20200722.

\subsection{The ``Reverberation machine" methodology} \label{method}

To maximize the statistical quality of the lag spectrum, data need to be grouped. As the evolution of BHLMXBs happens on typical timescales of days to months, the data grouping needs some criteria to group only the data with similar spectral-timing properties. Therefore, we start from a \nicer\ pilot pipeline (Steiner et al. in preparation) which processes public \nicer\ data and extracts the data products for each good time interval (GTI), including the flux-energy spectrum (both source and background), the light curves (LCs) and the power spectral densities (PSDs) in several different energy bands. 

The GTI-based library of data products enables efficient data grouping. We group data such that the relative changes compared to the first GTI in each group in the hardness ratio, the count rate, and the fractional variance ($F_{\rm var}$) are within a fixed fraction (by default, 30\%, 30\%, and 50\% respectively, but the exact values used for each source in our analysis are quoted in Table \ref{tab:sample_gg}); and the remaining detector combination after filtering out the noisy and hot detectors {\color{black}needs} to be the same. Specifically, the hardness ratio is defined as the count ratio of the hard band (4--12~keV) and the soft band (2--4~keV), and the count rate is in 0.3--12~keV. The fractional variance is calculated in 0.3--12~keV and within 1--10~Hz, the range where reverberation lags are expected for BHLMXBs. % and normalized for 52 FPMs

\begin{figure*}
\centering
\includegraphics[width=1.\linewidth]{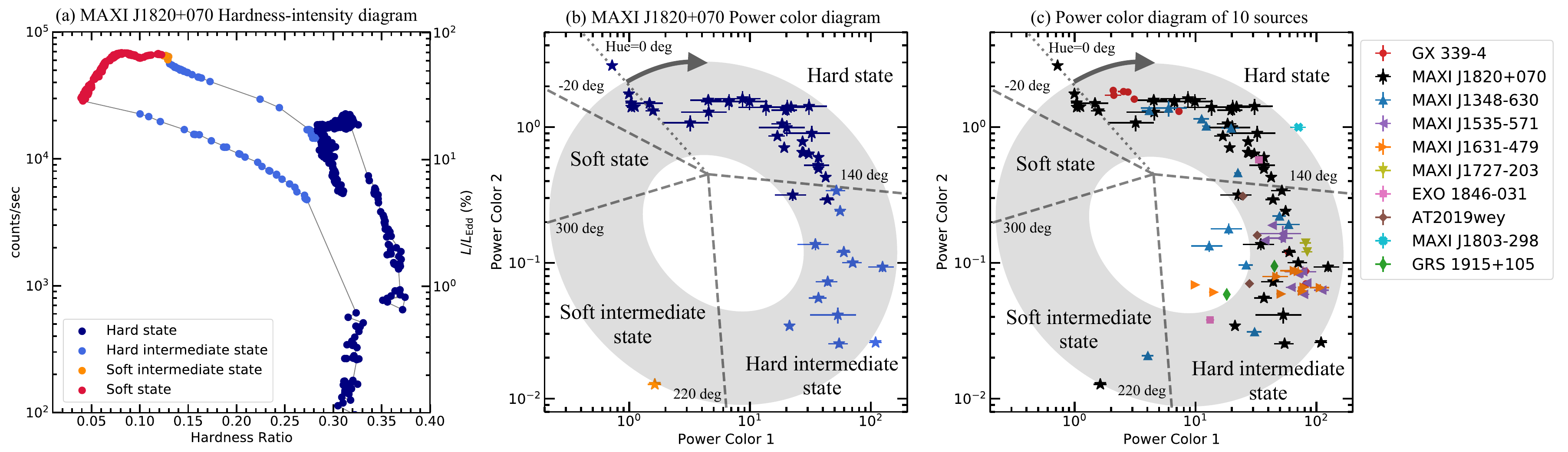}
\caption{We classify the spectral states and systematically compare the reverberation lag properties using the power-spectral ``hue" in the power color diagram. (\textit{a}) The hardness-intensity diagram of \maxij1820, for which the MJD corresponding to different spectral states are known from previous spectral-timing analysis \citep{buisson2019maxi,homan2020rapid,buisson2021maxi}. (\textit{b}) The power color diagram of \maxij1820\ in \nicer's 4.8--9.6~keV band. The hue ranges defined in \rxte's 2--13~keV band (boundaries between states shown in grey dashed lines) could distinguish the hard and hard intermediate states well. (\textit{c}) The power color diagram for the data groups of the 10 sources in which we detect soft lags. {\color{black}We note that the data points in the power color diagrams in (\textit{b}) and (\textit{c}) represent only those data groups with soft lags detected, and there is no good data group or soft lag detection in the soft state. }
}
\label{fig:hid_hue}
\end{figure*}

As the \nicer\ BHLMXB archive is large, and lag measurements for each data group would be time-consuming, we introduce a ``lag significance indicator" to filter only groups which are promising to generate good quality lag spectra, i.e.{\color{black}, those that} will result in small lag uncertainties. The phase lag uncertainty is

\begin{equation}
\footnotesize
\begin{multlined}
	\Delta\phi(\nu_j)=\sqrt{\left(\frac{P_{X,\rm noise}}{P_{X,\rm signal}}+\frac{P_{Y,\rm noise}}{P_{Y,\rm signal}}+\frac{P_{X,\rm noise}P_{Y,\rm noise}}{P_{X,\rm signal}P_{Y,\rm signal}}\right)/2M},
	\nonumber
\end{multlined}
\end{equation}
where $X$, $Y$ are the two energy bands between which lag is measured, ``noise" denotes the Poisson noise level, and ``signal" denotes the {\color{black}Poisson-noise-subtracted} intrinsic PSD, and $M$ is the number of LC segments. The last term dominates over the former two terms in the reverberation-dominated frequencies of BHLMXBs \citep{uttley2014x}. Therefore, if the intrinsic lag amplitude is similar and $P_{X,\rm signal}/P_{X,\rm noise}\propto P_{Y,\rm signal}/P_{Y,\rm noise}$, the signal-to-noise ratio (SNR) of a lag measurement is

\begin{equation}
\begin{multlined}
	{\rm SNR}_{\Delta t}\propto\sqrt{M}\cdot P_{\rm signal}/P_{\rm noise},
	\nonumber
\end{multlined}
\end{equation}
where $P_{\rm signal}$ and $P_{\rm noise}$ are the intrinsic PSD and Poisson noise in one energy band. Using the ``fractional rms-squared" normalization, the integrated intrinsic PSD over a frequency range is the fractional variance $F_{\rm var}$ due to variability in that range and the above estimate becomes

\begin{equation}
\begin{multlined}
	{\rm SNR}_{\Delta t}\propto\sqrt{M}\cdot F_{\rm var}\cdot \left<x\right>,
	\nonumber
\end{multlined}
\end{equation}
where $\langle x \rangle$ is the mean count rate. We therefore define the lag significance indicator as

\begin{equation}
\begin{multlined}
	{\rm SNR}_{\Delta t, ind}\doteq \sqrt{M}\cdot F_{\rm var}\cdot \left<x\right>,
	\nonumber
\end{multlined}
\end{equation}
which serves as a quantitative measurement of the expected quality of the lag spectrum. The longer the exposure, the more variable the LC is, the brighter the source gets, the larger the lag significance indicator ${\rm SNR}_{\Delta t, ind}$ is, resulting in a larger SNR of the lag measurement. For instance, a data group of a fairly bright BHLMXB (count rate $\sim$200~counts/sec) with total exposure of 10~ks ($M=1000$ when segment length is 10~s) and a typical fractional rms of 20\% in the hard state ($F_{\rm var}=0.04$) will have ${\rm SNR}_{\Delta t, ind}\sim250$. In the bright soft state, the fractional rms drops sharply to $\sim0.5\%$, so even if the count rate is as high as 10,000~counts/sec with a very long total exposure of 100~ks, ${\rm SNR}_{\Delta t, ind}\sim25$. 

From our experience working on previous datasets of \gx339\ \citep{wang2020relativistic} and \maxij1820\ \citep{wang2021disk}, we find roughly that an indicator value of $\sim250$ leads to a good-quality lag-frequency spectrum. This is not a strict limit, but rather a tool to help automate the pipeline. Therefore, for each data group, we calculate ${\rm SNR}_{\Delta t, ind}$ and filter only the ``good groups" reaching ${\rm SNR}_{\Delta t, ind}>250$. In three sources (MAXI~J1803--298, MAXI~J1727--203, and EXO~1846--031), there was no good group with the default grouping criteria. By relaxing the criteria on the relative changes in hardness ratio, fractional variance, and the count rate, there were a few groups that had indicator values slightly less than 250. We checked the lag-frequency spectra of these sources by hand, and determined that significant soft lags were present despite the slightly lower indicator value, and so we also include these in the sample (see Table~\ref{tab:sample_gg} for the data grouping criteria used after trials).

We classify the spectral state using the ``power-color" (PC) methodology of \citet{heil2015power}. PCs enable the shape of power spectrum to be characterised regardless of its normalisation in the same manner that spectral colors characterise the shape of flux-energy spectrum. We define PC1 as the integrated variance between 0.25--2.0~Hz divided by that between 0.0039--0.031~Hz, and PC2 is defined identically for the 0.031--0.25~Hz and 2.0-16.0~Hz bands. Fig.~\ref{fig:hid_hue}~(b) and (c) shows the power-color diagrams (PC2 vs PC1) of \maxij1820\ and all the 10 sources for which we detect soft lag (see Section~\ref{results:detections} for more details). We see, in line with the results of \citet{heil2015power}, that each outburst follows a circular path on the PC diagram and crucially, each source follows the same path. This is known as the PC ``wheel", and provides a very efficient way to define spectral states during an outburst. {\color{black}In a full outburst, the source follows the PC wheel in the \textit{clockwise} direction in the hard-to-soft state transition (i.e., from the hard state, through the HIMS and SIMS, to the soft state), and follows the same path but in the \textit{counter-clockwise} direction in the soft-to-hard state transition (i.e., from the soft state, through the SIMS and HIMS, to the hard state).} In particular, the \textit{hue} is defined as the clockwise angle from the axis defined at a 45$^\circ$ angle to the x and y-axes. The center of wheel is chosen to be (PC1, PC2)=(4.5, 0.45). The hue ranges for different spectral states are: hard state with hue from --20$^\circ$ to 140$^\circ$, HIMS with hue from 140$^\circ$ to 220$^\circ$, SIMS with hue from 220$^\circ$ to 300$^\circ$, and the soft state with hue from 300$^\circ$ to 20$^\circ$ (see Table 2 in \citealp{heil2015power}). We note that the original investigation of hue uses the \rxte\ data in 2--13~keV, and in our analysis, we use \nicer\ data in 4.8--9.6~keV to match the hard X-ray bandpass. We checked the data for \maxij1820\, for which the MJD corresponding to the spectral states are known from previous spectral-timing analysis \citep{buisson2019maxi,homan2020rapid,buisson2021maxi}, and found that the hue defined in this manner is indeed able to distinguish between hard and hard intermediate states (see Fig.~\ref{fig:hid_hue}~a--b).

For the good groups, we calculate the PSDs, the power-spectral hue, {\color{black}and} the lag-frequency spectrum measured between 0.5--1~keV and 2--5~keV. These two energy bands are chosen such that the soft band shows the soft excess, and the hard band is dominated by the coronal emission. We take light curve segment lengths of 10~s for the lag-frequency spectrum and 256~s for the PSD (to match the definition of hue), both with 0.001~s bins. If none of the GTIs in the group is longer than 256~s, the segment length for the PSD is also chosen to be 10~s; in this case, we still count the group as a good group if it meets the before-mentioned criteria because the segment length for the PSD does not affect the lag spectrum, and the only difference lies in the lack of a hue measurement. The logarithmic frequency rebinning factor is 0.4 for the lag-frequency spectrum, and 0.1 for the PSD. {\color{black}The former is chosen to increase the SNR of a lag measurement, and the latter is to to measure the QPO frequency more precisely. These choices do not change our results.} Both the source and background flux-energy spectra for each GTI in the data group are combined. The energy resolution of the source spectrum is oversampled by a factor of 3, and is then binned with a minimum count of 1 per channel. The response matrices for different data groups are combined according to the detectors used.

To keep the reverberation machine as automated as possible, we define criteria for determining if reverberation is present in the lag-frequency spectrum. At its most basic, we determine that reverberation is present if a significant soft lag (i.e.{\color{black},} a negative lag) is seen in the lag-frequency spectrum. In detail, reverberation is present if at least two frequency bins show a negative lag (i.e.{\color{black},} a soft lag) with its absolute amplitude larger than its uncertainty ($>2\sigma$ confidence equivalent).

It is further necessary to determine the frequency range over which a soft reverberation lag is present, and that task is performed as follows. As sometimes low statistics can lead to some points having lags consistent with zero, to avoid fragmenting the reverberation frequency range, we require that the range starts from a frequency bin with a negative lag, and continues to higher frequencies as long as the associated lag is consistent with being negative within $1\sigma$ confidence. This means the range ends when a positive (hard) lag appears with its amplitude larger than the uncertainty ($\nu_1$). In addition, in order to avoid biases due to phase wrapping, we define a ``phase wrapping frequency cut-off" ($\nu_2$) as the frequency when the absolute lag amplitude becomes more than half of the time lag corresponding to the phase limit $\pm\pi$. The reverberation lag range ends at either $\nu_1$ or $\nu_2$, whichever frequency is lower. 

The lags at the QPO frequencies can behave differently from their neighboring frequencies dominated by broadband noise, likely due to a different mechanism producing the QPOs \citep{van2016inclination,zhang2020systematic}. Therefore, as in our earlier work \citep{wang2021disk}, we fit the PSD with the sum of several Lorentzians, and calculate the lag-energy spectra outside of the central frequency $\pm$ half-width at half-maximum (HWHM) of those fitted QPOs\footnote{We choose the fundamental QPO to be the component with the largest rms and a quality factor $Q=\nu_0/{\rm FWHM}>2$.} and their (sub)harmonics. The reference band for the lag-energy spectra is the full band 0.3--10~keV.

All spectral fitting is done with XSPEC 12.11.1 \citep{arn96}. We use the PG-statistics, for Poisson data with a Gaussian background. We use the \textit{wilm} set of abundances \citep{wilms2000}, and \textit{vern} photoelectric cross sections \citep{verner1996atomic}, {\color{black}to account for the galactic absorption with the model \texttt{tbabs}.}

\begin{figure*}%[htb!]
\centering
\includegraphics[width=1.\linewidth]{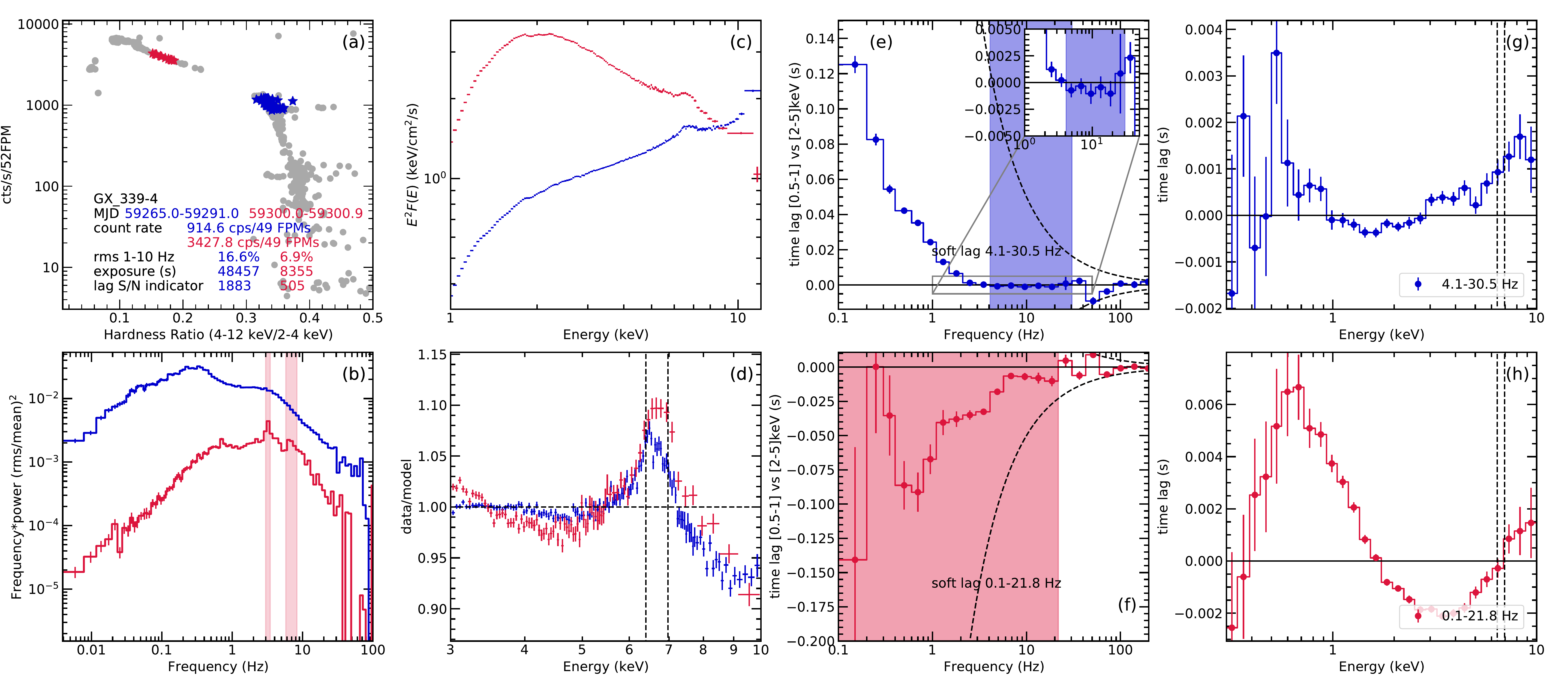}
\caption{Reverberation machine summary plot for \gx339\ (see Section~\ref{results:detections} for more details). Representative data groups in the bright hard state (blue) and during the hard-to-soft state transition (red). (\textit{a}) The GTI-based \nicer\ hardness-intensity diagram (HID) where the hardness ratio is defined as the count {\color{black}rate} ratio of the hard band (4--12~keV) and the soft band (2--4~keV), and the count rate is in 0.3--12~keV and normalized for 52 FPMs. Each dot represents a single GTI in the data group. (\textit{b}) The 0.3--10~keV PSDs with the shaded bands indicating the frequencies of the QPO and (sub)harmonics (central frequency$\pm$HWHM). (\textit{c}) The unfolded flux-energy spectra to a constant model. (\textit{d}) The data-to-model ratio when the {\color{black}entire} 3--10~keV flux-energy spectra are fitted with a \texttt{(diskbb+powerlaw)} model. The two dashed lines correspond to the energies of Fe K$\alpha$ at 6.4~keV and H-like Fe XXVI at 6.97~keV. (\textit{e})--(\textit{f}) The lag-frequency spectra measured between 0.5--1~keV and 2--5~keV. Shaded regions are where a soft reverberation lag is detected. The dashed lines represent the time lags corresponding to phase lags of $\pm\pi$. (\textit{g})--(\textit{h}) The lag-energy spectra in the frequency ranges where we detect soft lags. The two dashed lines are the same as in (\textit{d}), corresponding to the energies of Fe K$\alpha$ at 6.4~keV and H-like Fe XXVI at 6.97~keV. }
\label{fig:gx339}
\end{figure*}

\section{Results} \label{results}

\subsection{Reverberation lag detections}\label{results:detections}
Following the procedures in Section~\ref{method}, we run our autonomous reverberation search algorithm (``reverberation machine") over the sample. We find 10 sources with good data groups (see Table~\ref{tab:sample_gg}), and 16 sources with no good data group (see Table~\ref{tab:sample_no_gg}). Moreover, all the 10 sources with good groups have soft lag detections (see the detection criteria in Section~\ref{method}). For each source with a reverberation lag detection, we provide a brief introduction to the source properties, the outburst timeline, and the reverberation machine results including the number of good groups, the states covered, and the reverberation lag detections. We also present a summary figure that demonstrates what the data look like in the luminous hard state, and in the state transition (if available). As a case study, we show {\color{black}in this section} the results for \gx339\ which recently went on its first full outburst since \nicer's launch. {\color{black}The} results for the other 9 sources are detailed in the Appendix.

\gx339\ is an archetypal BHLMXB which outbursts every 2--3 years (e.g., \citealp{belloni2005evolution,motta2009evolution}; see also Fig.~1 in \citealp{garcia20192017}). The mass of the BH is constrained to be between 2.3 and 9.5~$M_\odot$, and our distance towards the system is $d>5$~kpc \citep{heida2017potential}. It is a mildly absorbed system with the column density of $N_{\rm H}\sim6\times10^{21}$~cm$^{-2}$ \citep{zdziarski1998broad}. The inclination via reflection spectroscopy is estimated to lie in the range of 30--48 degrees (e.g., \citealp{javier_gx339,wang2018evolution}), which makes it a famous low-inclination source. \gx339\ is significant because it was the first to show evidence for a reverberation feature in the hard state with \xmm\ \citep{uttley2011causal,de2015tracing,de2017evolution}. In the previous two hard-only outbursts in 2017 and 2019, \citet{wang2020relativistic} detected, for the first time with \nicer, a soft reverberation lag with amplitude of $9\pm3$~ms in the Fourier frequency range 2--7~Hz, in the brightest epoch available at that time. 

In January 2021, \gx339 underwent its first full outburst in the \nicer\ era (starting from obsID~3133010101 on 2021-01-20) and started the hard-to-soft state transition some time between 2021-03-20 and 03-26 (MJD 59293 and 59299, corresponding to obsID~4133010102 and 4133010103; \citealp{wang2021nicer}). Following the procedures described in Section~\ref{method}, we identify 14 good groups, and a soft reverberation lag is detected in 9 of them, in both the hard and intermediate states (covering data until obsID~4133010105 on 2021-03-28). 
An example of the reverberation machine summary figure is shown in Fig.~\ref{fig:gx339} for \gx339\, which includes (a) the hardness-intensity diagram (HID) with a representative hard state group and an intermediate state group shown in blue and red, respectively, (b) the full-band PSDs and any possible QPO feature, (c) the unfolded flux-energy spectra to a constant model, (d) the data-to-model ratio when the 3--10~keV flux-energy spectra are fitted with a \texttt{(diskbb+powerlaw)} model to see phenomenologically the iron line profile, (e)--(f) the lag-frequency spectra, and (g)--(h) the lag-energy spectra in the frequency ranges we detect soft lags. 
We see that the intermediate state exhibits a much larger lag amplitude that dominates at a lower frequency compared with the hard state; similar to what was previously observed for \maxij1820 \citep{wang2021disk}. We note that for clarity in the summary plots, we show only the lag-energy spectra in the full Fourier frequency band in which we detect soft lags (i.e., not outside of the QPO frequencies), but an example of the lag-energy spectra in the discontinuous frequency bands outside of QPO frequencies are shown in Fig.~\ref{fig:gx339_qpo_treatment}. We also exclude the measured lag within QPO frequencies when we estimate the reverberation lag amplitude for each data group in Section~\ref{lag_amp}.

The non-detections in the other 5 good groups of \gx339 are due to tentative soft lags in the lag-frequency spectra that do not meet the $2\sigma$ confidence criteria. Examples of these low-significance lags can be found in \citet{wang2020relativistic}. We also note {\color{black}the detection of} a soft lag depends on the chosen energy bands between which we measure the lags. The analysis presented in this systematic study adopts a uniform choice of a hard band of 2--5~keV (with soft band 0.5--1~keV). We have also tested a hard band of 1.5--3~keV as used in \citet{wang2020relativistic,wang2021disk}, and in that case, the soft lags in the lag-frequency spectra become more significant for the groups in the hard state, and less significant in the intermediate state. Fundamentally, this difference results from the different shapes of the lag-energy spectra, i.e., the corresponding energy of the ``valley" in which the lag is negative is approximately 1.5--3~keV in the hard state, and becomes closer to 2--5~keV during state transition (see panels~\textit{g} and \textit{h} in Fig.~\ref{fig:gx339}).

\subsection{The evolution of reverberation lag properties throughout the outbursts} 
\label{lag_amp}

\begin{figure*}
\centering
\includegraphics[width=0.7\linewidth]{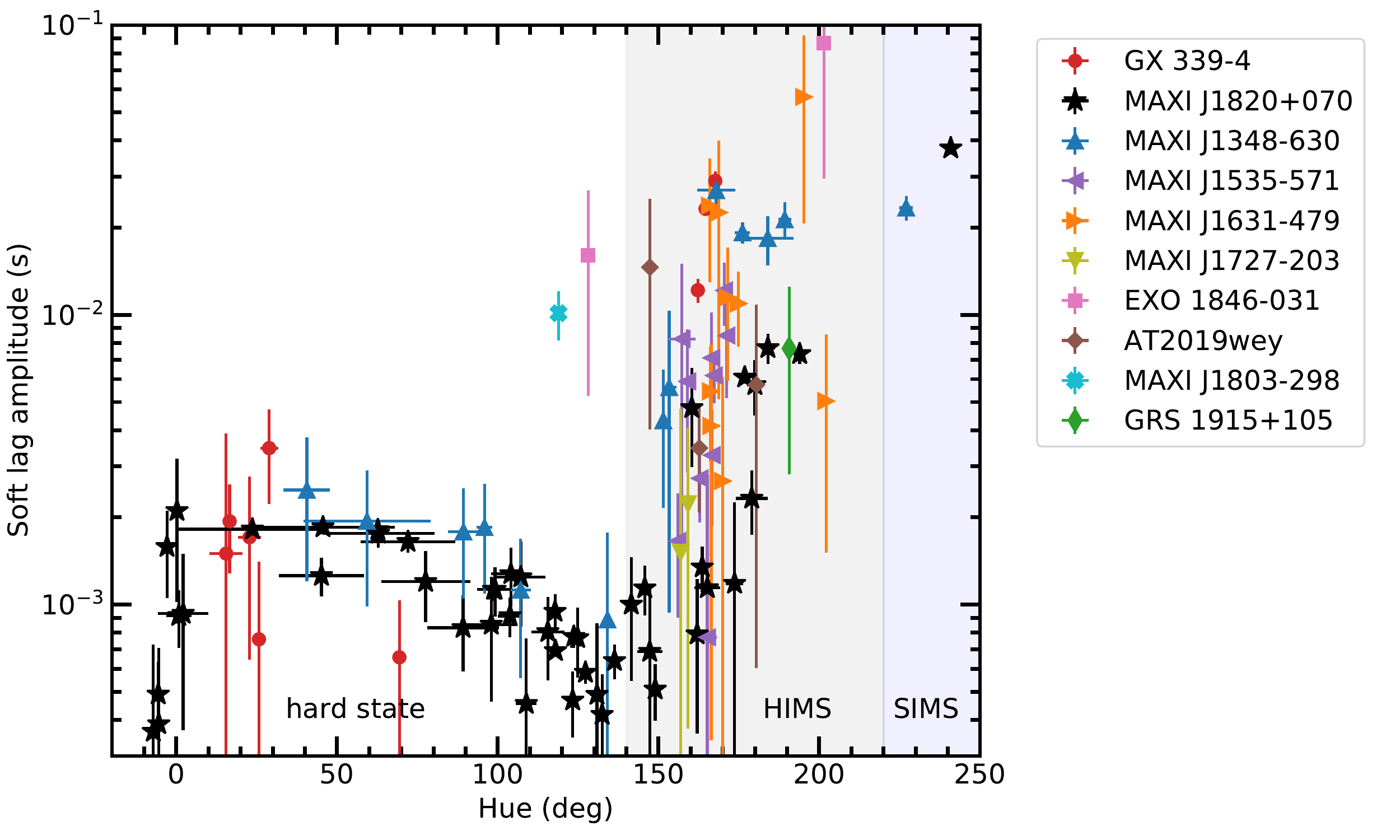}
\caption{A systematic look at the detected soft reverberation lags in the 10 BH/BHC LMXBs in which we detect significant soft lags. The evolution of soft lag amplitude with the power-spectral hue (see Sections~\ref{method} and \ref{lag_amp} for hue and lag amplitude estimates respectively). The reverberation lag becomes shorter in the hard state when hue$>20$ deg, and becomes longer after the state transition (HIMS shaded in grey and SIMS in blue). 
}
\label{fig:unbinned_hue_1}
\end{figure*}

\begin{figure*}
\centering
\includegraphics[width=0.9\linewidth]{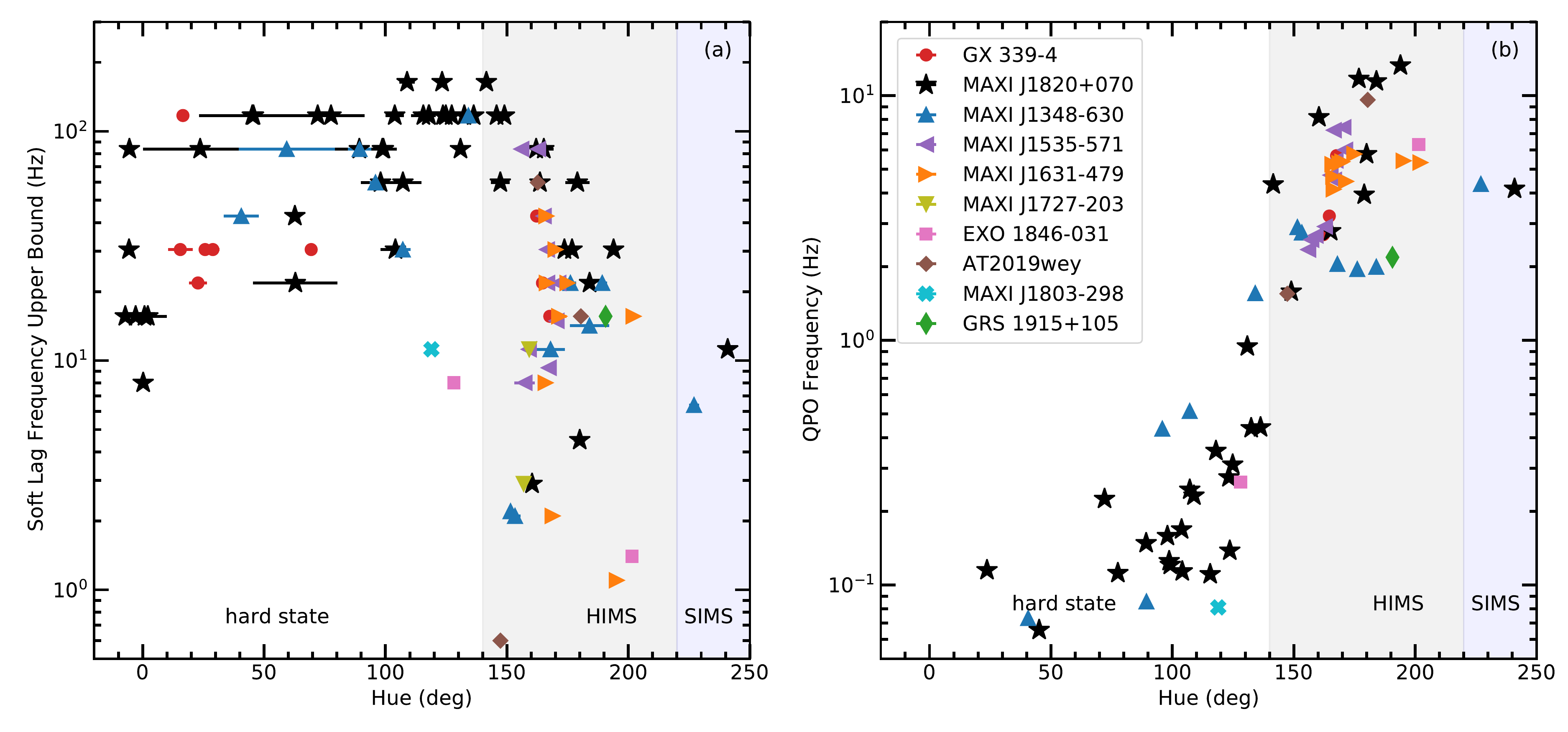}
\caption{A systematic look at the detected soft reverberation lags and the QPO frequencies in the 10 BH/BHC LMXBs in which we detect significant soft lags. (\textit{a}) The upper bound of the frequency band over which we observe a soft reverberation lag, and (\textit{b}) the QPO frequency. These two quantities both increase in the hard state (suggesting they are correlated), while they show opposite trends after state transition. 
}
\label{fig:unbinned_hue_2}
\end{figure*}

\begin{figure*}[htb!]
\centering
\includegraphics[width=1.\linewidth]{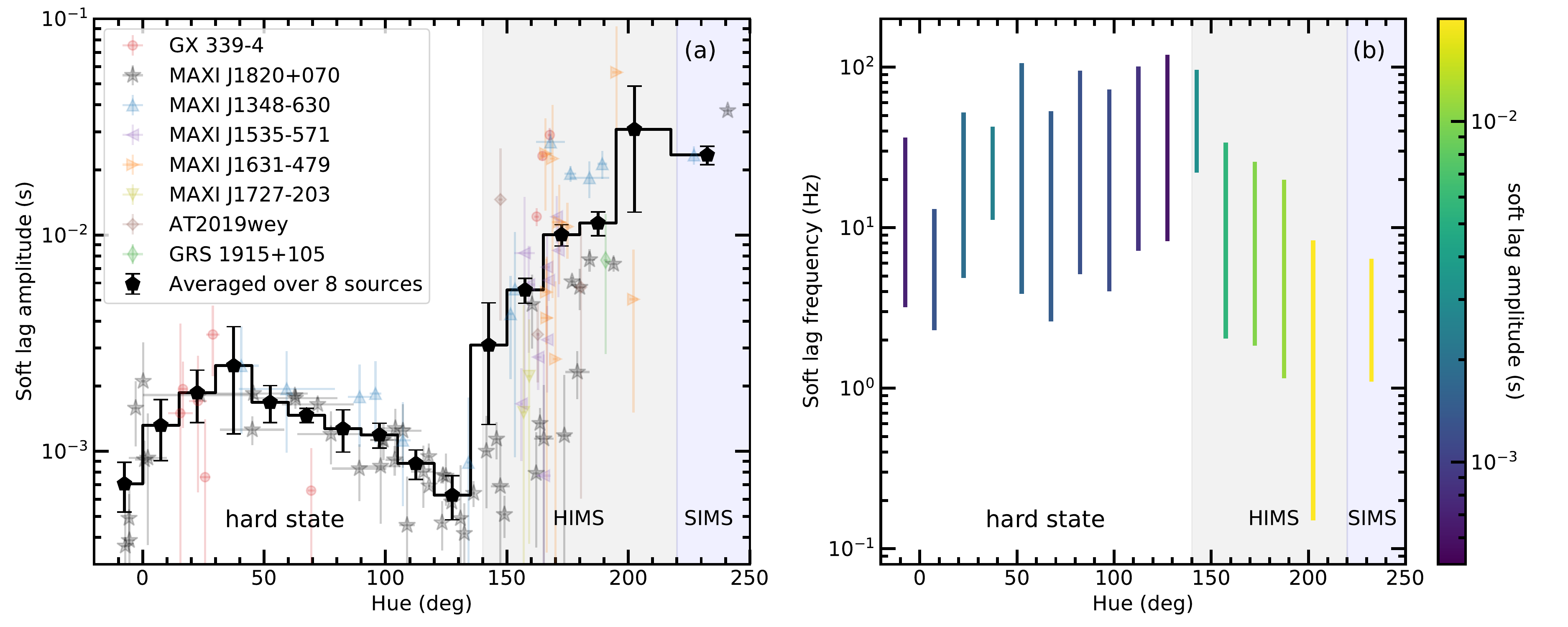}
\caption{(\textit{a}) The soft lag amplitude and (\textit{b}) the full Fourier frequency band where we detect soft lag rebinned by hue (i.e., averaged over continuous ranges of hue with widths equaling 15 degrees), for the 8 BH/BHC LMXBs excluding EXO~1846--031 and MAXI~J1803--298 (see discussion in Section~\ref{discussion:outliers}). The soft lag amplitudes for each individual source are also plotted in (\textit{a}). The colors matching the colorbar in (\textit{b}) indicate the soft lag amplitudes shown in (\textit{a}). The uncertainties in (\textit{a}) are the statistical errors of soft lag amplitudes in each range of hue. We also note that the vertical lines in (\textit{b}) represent the full Fourier frequency band where we detect soft lag, not the upper bound of that frequency band as shown in Fig.~\ref{fig:unbinned_hue_2}a. In the hard state, the reverberation occurs at increasing frequencies, with a decreasing lag amplitude; while after the state transition (HIMS is shaded in grey and SIMS in blue), the reverberation happens at much lower frequencies, with a much larger lag amplitude. 
}
\label{fig:6sources_binned}
\end{figure*}

With reverberation lags detected in 10 different BH/BHC LMXBs (see Table~\ref{tab:sample_gg}), we can perform a systematic comparison of the reverberation lag properties from source to source, which could help us to understand the disk-corona geometry across spectral states.

As mentioned in Section~\ref{method}, we use the power-spectral hue as an indicator of the state the sources are in during their outbursts (see Section~\ref{method}). The reverberation lag amplitude is approximated with a variance-weighted average of the soft lag amplitude in the lag-frequency spectrum (i.e., the soft lag amplitude is weighted based on its uncertainty $\sigma$ in each frequency bin we detect a soft lag, by a factor of $1/\sigma^2$). We exclude a frequency bin if its centroid lies within the QPO (and their harmonics or sub-harmonics) centroid frequency $\pm$ HWHM.

The resulting soft lag amplitude of the 10 BH/BHC LMXBs as a function of the hue is shown in Fig.~\ref{fig:unbinned_hue_1}. Our 10 sources show a remarkably similar reverberation lag behaviour as a function of spectral state. We discuss this in more detail later in Section~\ref{discussion:lag_amp}. As the averaged lag amplitude depends on the level of dilution, a useful quantity to measure the reverberation lag properties is the upper bound of the frequency band where we detect a soft lag, which should not be affected by dilution (while the lower bound can be affected by a change in the low-frequency hard lags). The evolution of the upper bound of the soft lag frequency band across values of hue is shown in Fig.~\ref{fig:unbinned_hue_2}a, together with the fundamental QPO centroid frequency in Fig.~\ref{fig:unbinned_hue_2}b. 
Except for two outliers in the hard state (EXO~1846--031 and MAXI~J1803--298, see Section~\ref{discussion:outliers} for the discussion), all sources appear to follow the same evolution of soft lag amplitude with hue. We therefore average over all sources except for the two outliers by binning on hue. The resulting rebinned evolution of the frequency and amplitude of the soft lag as functions of hue for the remaining 8 sources are shown in Fig.~\ref{fig:6sources_binned}, and the soft lag amplitude as a function of QPO frequency is shown in Fig.~\ref{fig:6sources_binned_lag_qpo}.
We summarize the findings as follows: 

\begin{figure*}
\centering
\hspace*{-1cm}
\includegraphics[width=0.6\linewidth]{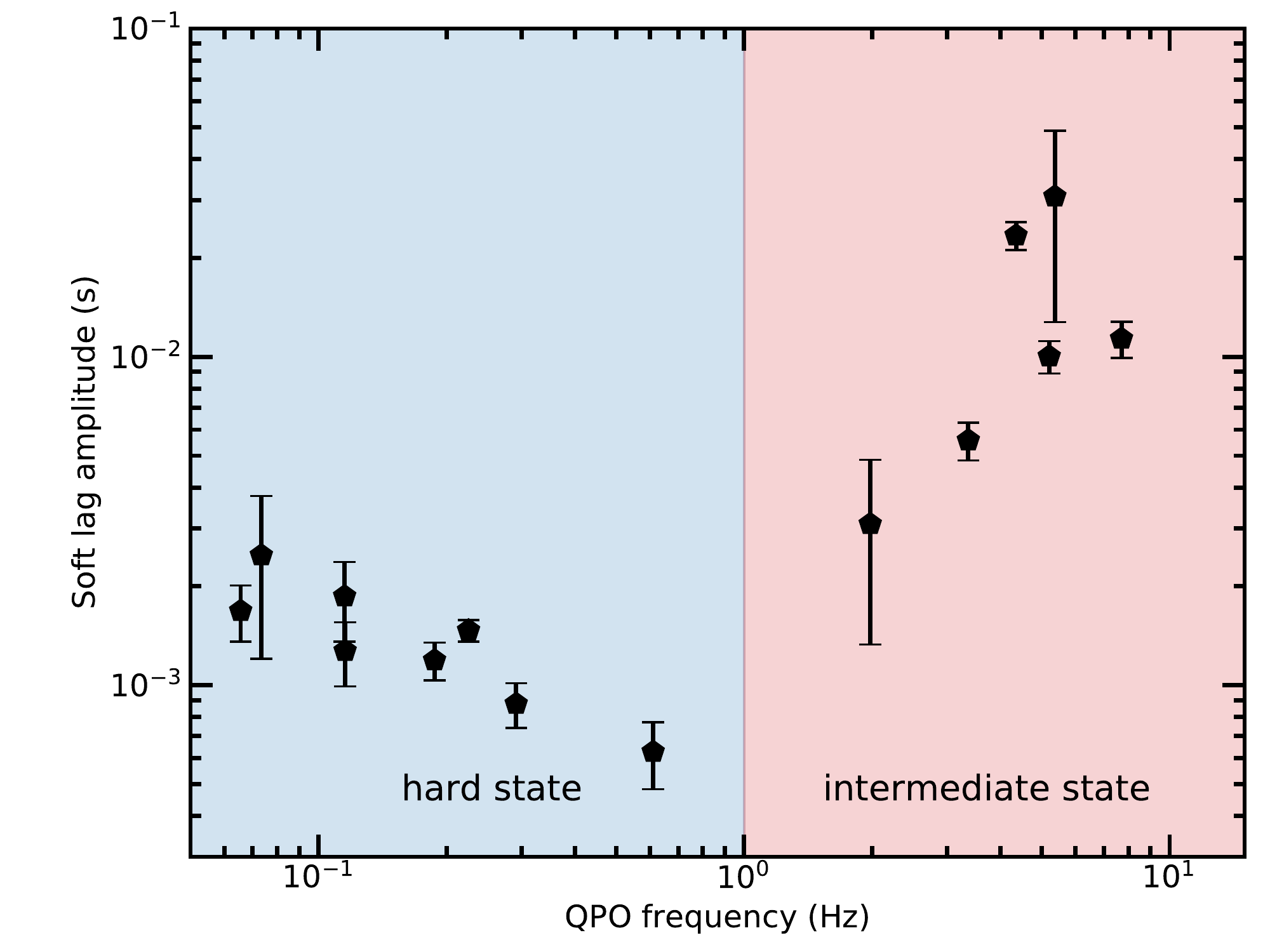}
\caption{The soft lag amplitude versus the QPO centroid frequency rebinned by hue. In the hard state, the QPO frequency is $<1$~Hz, and the soft lag amplitude decreases with QPO frequency, as expected if QPO frequency and the frequency where reverberation is detected are correlated. However, QPO and reverberation become anticorrelated in the intermediate state. The uncertainties are the statistical errors of soft lag amplitudes in each range of hue. See Section~\ref{discussion:qpo} for the implication. 
}
\label{fig:6sources_binned_lag_qpo}
\end{figure*}

\begin{itemize}
    \item In the hard state (hue $<140^\circ$), as hue increases, we see an increase in soft lag amplitude at the beginning (when hue is $\lesssim20^\circ$, see Section~\ref{discussion:hard_state} for the discussion), and then all the 3 sources with more than 1 reverberation lag detection in this state (\gx339, \maxij1820, and MAXI~J1348--630) agree on a steadily decreasing trend of lag amplitude ($\sim2$~ms to $\lesssim1$~ms, see Fig.~\ref{fig:unbinned_hue_1} and Fig.~\ref{fig:6sources_binned}a). Meanwhile, we see a tentative increase of soft lag frequency with hue from tens of Hz to $\sim100$~Hz (see Fig.~\ref{fig:unbinned_hue_2}a and Fig.~\ref{fig:6sources_binned}b). The evolution of both quantities indicates that the size of the region causing reverberation (i.e., the inner disk-corona region) 
    reduces as the hard state evolves. This is consistent with what has previously been reported for \maxij1820\ and interpreted as a contracting corona in the luminous hard state \citep{kara2019corona,buisson2019maxi}. The QPO frequency increases from $\sim0.1$~Hz to $\sim1$~Hz (Fig.~\ref{fig:unbinned_hue_2}b). This can explain why both the Type-C QPO and reverberation evolve to shorter timescales (higher frequencies) before the transition to HIMS. The correlation between QPO and reverberation can also be seen in Fig.~\ref{fig:6sources_binned_lag_qpo} as long as the QPO frequency is $<1$~Hz, as a higher upper bound of the frequency band over which reverberation occurs generally comes with a shorter lag.

\begin{figure*}
\centering
\includegraphics[width=1.\linewidth]{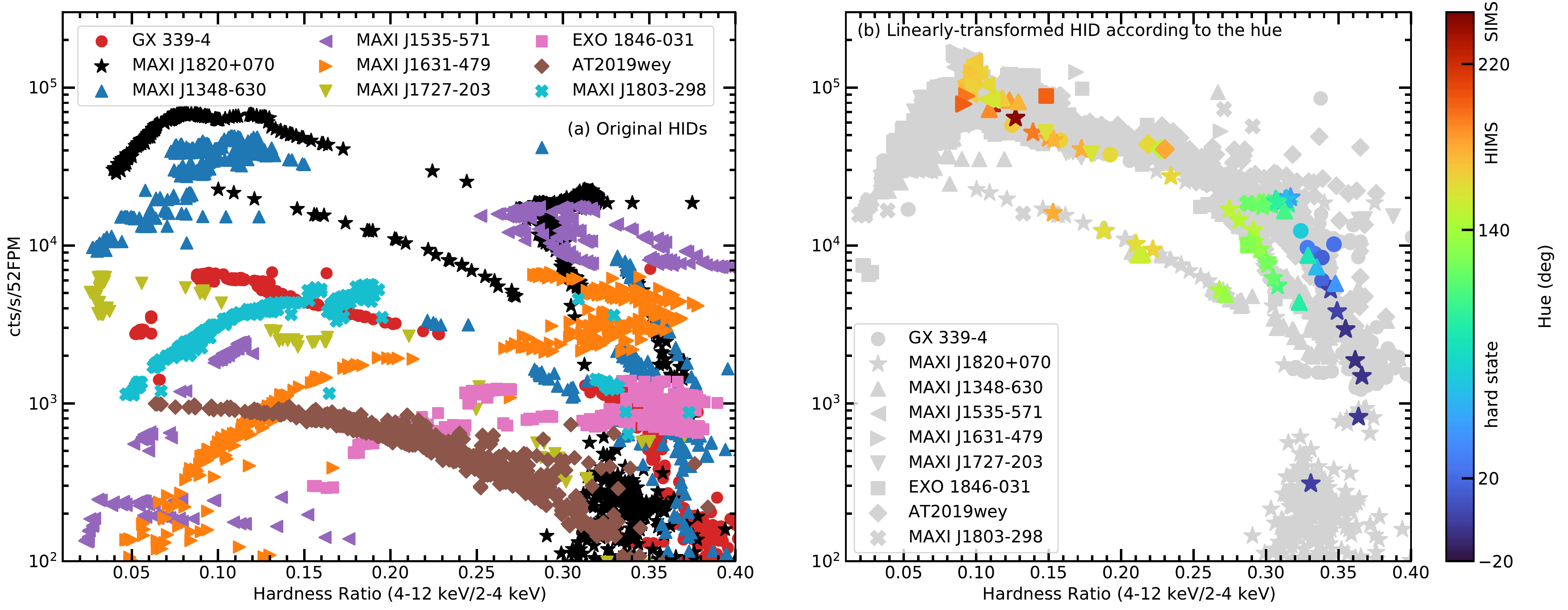}
\caption{The \nicer\ HIDs for the BH/BHC LMXBs in which we detect soft reverberation lags (GRS~1915+105 is not included because of its unconventional spectral states). (\textit{a}) The original HIDs, in which the high galactic absorption of MAXI~J1535--571 (purple), MAXI~J1631--479 (orange), and EXO~1846--031 (pink) make their HIMS appear in the same hardness as the hard state in other sources. (\textit{b}) The combined HID (grey) after making linear transformations for both axes in the 8 sources except for the benchmark \maxij1820. The transformation aims at making observations with similar hues lie in the same region of the HID. The values of hue for the data groups with soft reverberation lag detected are shown with colors matching the colorbar. 
}
\label{fig:unbinned_hid_1}
\end{figure*}
    \item In the HIMS ($140^\circ<$ hue $<220^\circ$), as the hue increases, the reverberation lag amplitude shows a sharp increase by a factor of 10, from $\lesssim1$ to $\sim10$~ms (see Fig.~\ref{fig:unbinned_hue_1} and Fig.~\ref{fig:6sources_binned}a). Of the 9 sources with more than 1 reverberation detection, all agree with this increase. In EXO~1846--031, the only hard state detection has an anomalously high lag amplitude and correspondingly low upper bound frequency for which reverberation dominates (see Section~\ref{discussion:outliers} for the discussion). The reverberation lag amplitudes are mostly consistent (within uncertainties) with \maxij1820, which has the best coverage and the most detections. At the same time, the upper bound of the frequency where reverberation is observed drops from $\sim100$~Hz at the end of hard state to $\sim10$~Hz (see Fig.~\ref{fig:unbinned_hue_2}a). In Fig.~\ref{fig:6sources_binned}~(b), we see it is not only the upper bound of the reverberation lag frequency that decreases, but also the lower bound, i.e., we could safely say that the reverberation signal evolves to lower frequencies. This result agrees with our \maxij1820\ result \citep{wang2021disk} and provides further evidence for a longer reverberation lag in the HIMS, and is consistent with a growing disk-corona distance, possibly due to a vertically expanding corona sitting at the base of a jet. By comparing the timescale of the reverberation lag to that of the Type-C QPO, we see that unlike in the hard state, during the state transition, the Type-C QPOs continue to evolve to higher frequencies (from $\sim1$~Hz to $\sim10$~Hz), but the reverberation lags, instead, get longer (see Fig.~\ref{fig:6sources_binned_lag_qpo} when the QPO frequency is $>1$~Hz). We discuss implications of these trends between QPO and reverberation in Section~\ref{discussion:qpo}. 
    \item In the SIMS (hue $>220^\circ$), out of the 10 sources in our sample, we have only 2 good groups (in \maxij1820\ and MAXI~J1348--630) because of the sudden decrease in the variability level compared to the HIMS. A reverberation lag is detected in both data groups, and the mean lag amplitude is even larger ($\sim0.03$~s) compared to the HIMS. 
\end{itemize}

\begin{figure*}[htb!]
\centering
\includegraphics[width=1.\linewidth]{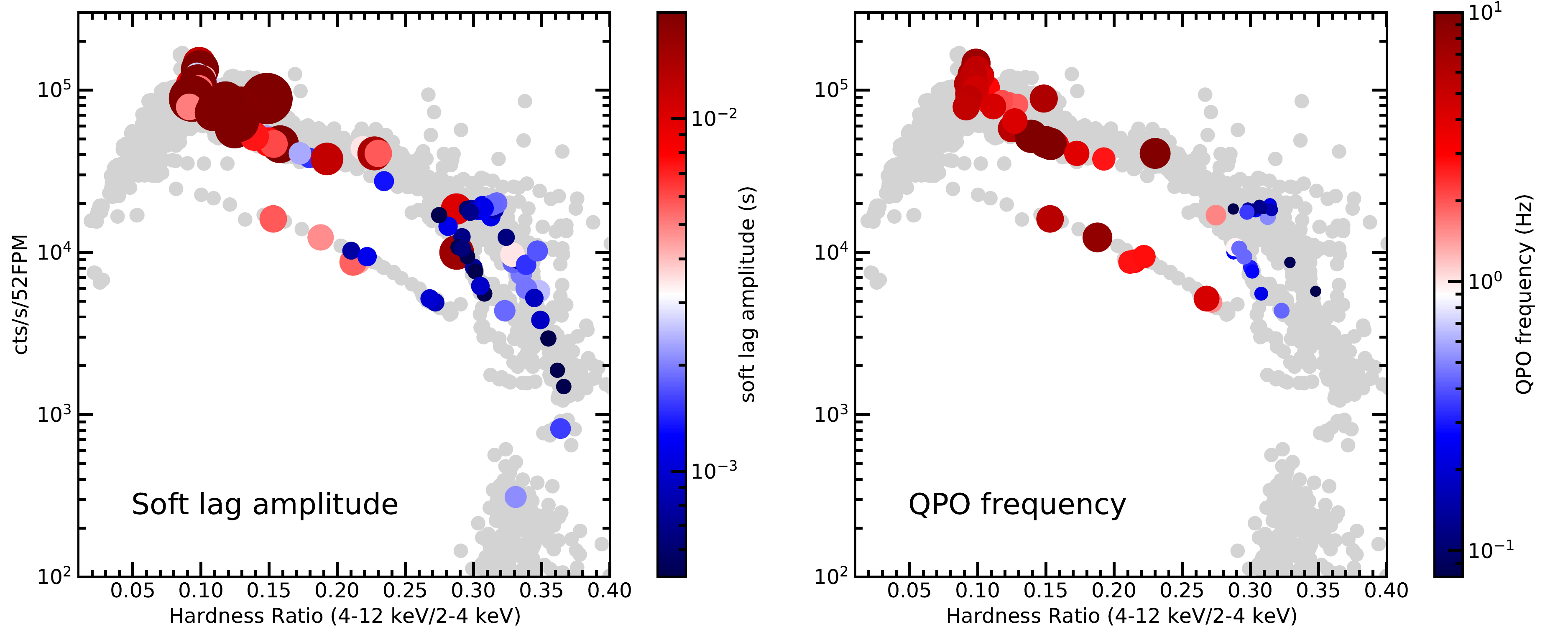}
\caption{The (\textit{left}) soft lag amplitude and (\textit{right}) the QPO centroid frequency in the combined HID (grey). During the state transition, the reverberation lag amplitude increases from $\sim1$~ms to $\sim10$~ms. The QPO frequency increases in the hard state and the HIMS. Notice the values of soft lag amplitudes and QPO frequencies are the same as those shown in Figs.~\ref{fig:unbinned_hue_1} and \ref{fig:unbinned_hue_2}b, and are represented by the colors matching the colorbars. {\color{black}In addition to the different colors matching the colorbars, the marker sizes also grow with the soft lag amplitude and QPO frequency.} See Fig.~\ref{fig:unbinned_hid_1} for more details on the combined HID. 
%The luminosity in Eddington units is obtained assuming the BH mass ($8.48^{+0.79}_{-0.72}$~$M_\odot$; \citealp{torres2020binary}) and distance (2.96$\pm$0.33~kpc) of \maxij1820\ as it is the benchmark whose HID is not linearly transformed.
}
\label{fig:unbinned_hid_2}
\end{figure*}

To compare the reverberation lag amplitude and the QPO frequency in a more intuitive way, we attempt to unify the HIDs for the 9 sources (except for GRS~1915+105, due to its nonstandard spectral evolution). Simply overplotting the raw HIDs (Fig.~\ref{fig:unbinned_hid_1}a) can be misleading because each source is at a different distance, with a different associated galactic absorption, etc. This means, for example, that a hardness of 0.3 for \maxij1820\ corresponds to its hard state (according to the power-spectral hue), but the same hardness for MAXI~J1535--571 corresponds to the HIMS because MAXI~J1535--571 is so heavily obscured. We attempt to scale all sources so that each state (defined by its hue) matches the HID of our best sampled source, \maxij1820. Five of the sources (\gx339, MAXI~J1348--630, MAXI~J1727--203, AT2019wey, and MAXI~J1803--298) mostly only require a shift in the count rate axis to match the benchmark (likely simply due to the fact that they are at different distances), but the other three sources (MAXI~J1535--571, MAXI~J1631--479, and EXO~1846--031) require both a shift in hardness ratio and count rate. Indeed, these three sources have larger Galactic column densities compared to the rest of the sample (in unit of $10^{22}$~cm$^{-2}$, $N_{\rm H}\sim5$ in MAXI~J1535--571, \citealp{xu2018reflection}; $N_{\rm H}\sim3$ in MAXI~J1631--479, \citealp{monageng2021radio}; $N_{\rm H}\sim11$ in EXO~1846--031, \citealp{draghis2020new}). The larger galactic absorption diminishes the count rate in the 2--4~keV band when calculating the hardness ratio, and therefore shifts the hardness ratio to higher values, which is exactly what we observe in Fig.~\ref{fig:unbinned_hid_1}~(a). 

In the 8 sources except for the benchmark \maxij1820, we make linear transformations for both axes, so that observations with similar hues lie in the same region of the HID. The result of this transformation can be seen in Fig.~\ref{fig:unbinned_hid_1} (b). The final result is shown in Fig.~\ref{fig:unbinned_hid_2}, where the soft lag amplitudes and the QPO frequencies are presented on the combined HID. One can clearly see that the reverberation lags become much longer in the state transition, while the QPO frequencies continue to increase, which is also shown in Fig.~\ref{fig:6sources_binned_lag_qpo}. We note that the linear transformation of HID is only for the purpose of presenting the reverberation lag amplitudes and QPO frequencies in a canonical HID. The values of lag amplitudes and QPO frequencies are independent of the linear transformation of HID.

\section{Discussion} \label{discussion}

With our reverberation machine designed for systematic reverberation lag searches, we have detected soft reverberation lags in 10 BH/BHC LMXBs in the \nicer\ archive. This work has not only increased the sample size by a factor of 5, but also expanded our horizons from the hard state to the intermediate state. We compare the evolution of the soft lag amplitudes in different outbursts in a systematic way, and find that the reverberation lags are short in the hard state (lag amplitudes of $\sim 1$~ms) and increase by a factor of $\sim 10$ during the state transition (e.g., see Fig.~\ref{fig:unbinned_hue_1}).

\subsection{The disk/corona/jet geometry in the state transition}\label{discussion:lag_amp}

In the most naive picture, the reverberation lag, $\tau$, relates to the distance between a lamppost corona and reflector, $d$, as $\tau = (d/c) [ 1 + \cos i]$ where $i$ is the inclination angle. Therefore, for a given intrinsic disk-corona geometry in units of $R_g$, a higher BH mass (i.e.{\color{black},} a larger $R_{\rm g}$) leads to a longer reverberation lag. Alternatively, for a given BH mass, the lag increases with increasing $d/R_g${\color{black}, which is related to intrinsic change of the disk-corona geometry in units of $R_g$. }

Among our sample, \maxij1820\ has the most reliable measurements of its BH mass ($8.48^{+0.79}_{-0.72}$~$M_\odot$; \citealp{torres2020binary}) and inclination (63$\pm$3 degrees; \citealp{bright2020extremely}). Assuming these values, a reverberation lag amplitude of 1~ms corresponds to a light travel time of $\sim16$~$R_{\rm g}/c$ in the lamppost geometry. Then, a decrease of reverberation lag amplitude from 2~ms to $1$~ms in the hard state corresponds to a decrease in the distance between the corona and the inner disk of $32$~$R_{\rm g}$ to 16~$R_{\rm g}$. For \maxij1820, we adopted the reverberation model \reltrans\ to fit the lag-energy spectra in several frequency bands, and measured the coronal height in the lamppost geometry to decrease from 40 to 27~$R_{\rm g}$ in the hard state \citep{wang2021disk}. {\color{black}The \reltrans\ model computes the reverberation lags from the light travel times of the photons using general relativistic ray tracing, and accounts self-consistently for the dilution. The estimate from the proper modeling 
%accounting self-consistently for the dilution and also general relativity
is close to the naive estimate, and the discrepancy within a factor of 2 is mostly due to dilution.}

\begin{figure*}
\centering
\includegraphics[width=1.\linewidth]{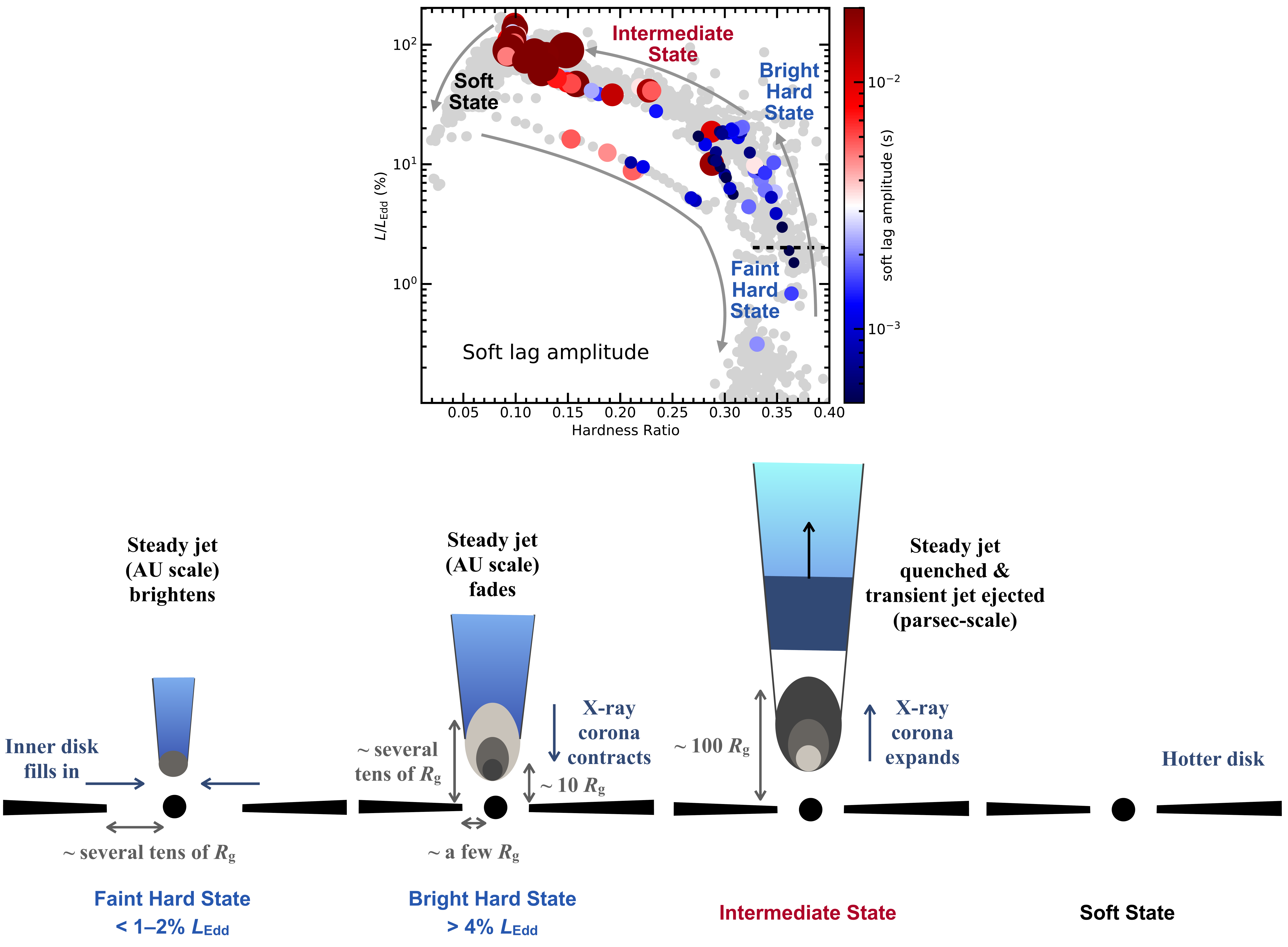}
\caption{{\color{black}(\textit{Upper}) The measured soft lag amplitudes in the combined linearly-transformed HID (same as Fig.~\ref{fig:unbinned_hid_2}a). The luminosity in Eddington units is obtained assuming the BH mass ($8.48^{+0.79}_{-0.72}$~$M_\odot$; \citealp{torres2020binary}) and distance (2.96$\pm$0.33~kpc) of \maxij1820\ as it is the benchmark whose HID is not linearly transformed. (\textit{Lower}) The evolution and geometry of the disk-corona-jet system we infer in the faint hard state, bright hard state, and the intermediate state. In the faint hard state ($<1-2\%$~$L_{\rm Edd}$), the disk is truncated at several tens of $R_{\rm g}$ and the inner disk fills in, causing the reverberation lag to become shorter (see Section~\ref{discussion:hard_state} and Fig.~\ref{fig:hard_state_lag_edd}). In the bright hard state ($>4\%$~$L_{\rm Edd}$), the disk is no longer truncated, and the X-ray corona vertically contracts from several tens of $R_{\rm g}$ to $\sim10$~$R_{\rm g}$ (see Section~\ref{discussion:lag_amp} and Fig.~\ref{fig:hard_state_lag_edd}), while the steady jet fades. In the intermediate state, the X-ray corona expands to $\sim100$~$R_{\rm g}$, while the steady jet is quenched and the transient jet is ejected (see the discussion in Section~\ref{discussion:lag_amp}). {\color{black}In the soft state, there is no good data group (due to the low variability) or soft lag detection, but we note that there is a great amount of evidence that the disk extends to the ISCO radius (e.g., \citealp{steiner2010constant}); the disk emission (with a disk temperature of $\sim1$~keV) dominates over the coronal emission \citep{done2007modelling}, and the radio jet is off \citep{fender2004towards}. During the soft-to-hard state transition and the following decay in the hard state, the soft lag amplitudes are consistent with those during the rise at earlier time in the outbursts. However, as the number of detection at this stage is limited, future data and more detailed modeling are needed to test for any hysteresis effect during the rise and decay in the hard state, or in the upper and lower branches of the intermediate state.}}
}
\label{fig:schematics}
\end{figure*}

Then, during the state transition, the increase of reverberation lag amplitude to 10~ms leads to a disk-corona distance of $\sim$140~$R_{\rm g}$. This is of the same order of magnitude as the fitted coronal height for \maxij1820\ with \reltrans\ (starting from $\sim100-200$~$R_{\rm g}$ and eventually to $\gtrsim300$~$R_{\rm g}$; \citealt{wang2021disk}). Moreover, since \maxij1820\ has the closest-in-time radio coverage to the X-ray state transition, we were able to {\color{black}determine} that the radio flare caused by the ballistic transient jet lagged behind the X-ray corona expansion by $\sim5$~days. 
%During the same state transition, a long soft lag was detected which hints on a re-filling disk $\sim1$ day after the jet ejection \citep{buisson2021maxi}. These findings suggest 
{\color{black}This finding suggests} that the state transition is marked by the corona expanding and ejecting a jet knot that propagates along the jet stream at relativistic velocities, supporting the case for the corona being the base of the jet {\color{black}(see the schematic referred to as ``intermediate state" in Fig.~\ref{fig:schematics})}.

Remarkably, we find that in all the 9 sources with more than 1 reverberation lag detection, as previously reported for \maxij1820\ \citep{wang2021disk}, the reverberation lag gets longer in the state transition. This result shows this trend is a generic feature of state transitions of BHLMXBs, not depending on, e.g., inclination, spin, or mass. Unfortunately, we cannot test for a correlation between lag amplitude and BH mass, inclination, or spin, as most of our sample consists of new transients lacking robust measurements of these quantities. {\color{black}For different BHLMXBs, the dependency of the lag amplitude on the BH mass and inclination can be seen from the naive picture at the beginning of this section. The lag amplitude could also depend on the BH spin because it could change the disk-corona distance in units of $R_g$ in two ways: (1) the innermost stable circular orbit (ISCO) radius monotonically decreases with the BH spin; (2) as the jet could be powered by the BH spin (see e.g., \citealp{narayan2012observational,steiner2012jet}), the geometry of the corona (as base of a jet) could thus be related to the BH spin.} In addition, the inclination could only cause a difference in lag amplitude by a factor of 2 in the naive picture, and the BH mass is usually the same within a factor of 2. That is to say, the reverberation lag amplitude is expected to have a narrow dynamic range, making it very challenging to test for its correlation with physical quantities.

The generic trend of reverberation lag getting longer is highly suggestive of what happens in such state transitions. The reverberation lag evolution during the state transition cannot result merely from dilution, because the upper bound of the Fourier frequency where reverberation is detected also evolves to lower values (see Fig.~\ref{fig:unbinned_hue_2}a and \ref{fig:6sources_binned}b). This upper frequency bound is independent of dilution. Therefore, a growing size of the disk-corona region that causes reverberation is necessary. In our modeling of \maxij1820\ data, we find an increase in the coronal height possibly due to a vertically expanding corona. 

We also note that a new time lag model for accreting BHs could explain the hard lags and soft (reverberation) lags with a relatively compact corona (Uttley \& Malzac 2021, in preparation; see \href{https://zenodo.org/record/5515847#.YVbKD9P7RBw}{the poster}). Uttley \& Malzac suggested that the power-law pivoting delay which produces the hard power-law continuum lags, modelled empirically in \reltrans, is due to the delay expected between seed photon variations and coronal heating as the driving mass accretion fluctuations propagate from the disk (providing the seed photons) to the corona (where coronal heating occurs). In this scenario, hard lags are produced at lower frequencies \textit{and} soft lags are produced at higher frequencies, due to accretion propagation effects. The soft lags result from the lag of the reverberation signal (which responds primarily to the {\color{black}power-law emission in the harder X-rays}, driven mainly by coronal heating) relative to the seed photons from fluctuations propagating through the disk (which cause {\color{black}the power-law flux in the soft band to rise before the power-law flux in the hard band}). However, these lags also depend on the coronal geometry, so that more vertically extended {\color{black}coronae} show larger soft lags, i.e.{\color{black},} the inferred geometric change is the same as for the reverberation lags being due to light-travel times only. However, in the case where the soft lags are linked to propagation, the vertical scales required to produce the lags are reduced.

The idea of the corona expanding in state transition was previously suggested on different grounds \citep{pottschmidt2000temporal,nowak2002coronal}. The authors used \rxte\ data of Cygnus X--1 and \gx339\, and found longer \textit{hard} lags during state transition. The interpretation was that these prolonged lags were produced in spatially very extended synchrotron-emitting outflows. More recently, this result was generalized into the photon-index-time-lag correlation in BHXBs (e.g., \citealp{grinberg2013long,reig2018photon}). With \nicer's soft X-ray bandpass and large effective area, we are now able to find prolonged \textit{soft reverberation} lags in state {\color{black}transitions}, which {\color{black}provide} more direct evidence of the expanding corona {\color{black}since} the mechanism that produces hard lags is not well understood. In addition to longer lags, recent broadband spectral modeling for several BHLMXBs finds that the radius of the corona which is assumed to be the base of the jet increases from $\sim10$~$R_{\rm g}$ to $\sim40$~$R_{\rm g}$ through the hard state and the HIMS; and the reflection fraction occasionally becomes very low in the HIMS when approaching the SIMS, suggesting an increase in also the vertical extent of the corona and/or its bulk velocity (\citealt{lucchini2021correlating,cao2021evidence}, but see \citealt{steiner2017self}).

\subsection{The QPO/reverberation connection}\label{discussion:qpo}

We compare the timescale of the reverberation lag to that of the Type-C QPO in Fig.~\ref{fig:unbinned_hue_2}. In the hard state, the reverberation lags evolve to shorter timescales (higher frequencies) before the transition. Similarly, the Type-C QPOs also evolve to higher frequencies. In fact, all characteristic frequencies in the power spectrum increase during the outburst, not just the QPO frequencies \citep{1999ApJ...514..939W,1999ApJ...520..262P}. However, during the state transition, the Type-C QPOs continue to evolve to higher frequencies, but the reverberation lags, instead, get longer (Fig.~\ref{fig:6sources_binned_lag_qpo}). In short, reverberation and Type-C QPO are correlated in the hard state, but they anticorrelate during the state transition. By modeling the reflection spectrum with a dual-lamppost model {\color{black}(two lampposts at different coronal heights) to mimic a vertically extended corona}, \citet{buisson2019maxi} also found the upper coronal height to decrease first in the bright hard state, and then increase when approaching the state transition in \maxij1820, while the characteristic features of the temporal variability kept increasing in frequency.

Type-C QPOs' amplitudes show great evidence for an inclination dependence (e.g., \citealp{motta2015geometrical,heil2015inclination}), which suggests that they arise from some geometric {\color{black}effect}, e.g.{\color{black},} Lense-Thirring precession of an inner hot flow or corona \citep{ingram2009low}. The evolution of the QPOs to higher frequencies has been proposed as evidence for a smaller inner hot flow or central corona (along with a less truncated disk) before the source reaches the soft state. This could also explain the increasing disk blackbody temperature, the softening coronal emission with a decreasing coronal temperature \citep{done2007modelling}, and also the shorter reverberation lags in the bright hard state \citep{kara2019corona}. However, the anticorrelation of the QPO frequency and reverberation lag in the state transition breaks this simple picture.

{\color{black}If both the QPOs and reverberation lags depend mostly on the geometry of the corona-inner-disk system as discussed before, t}he anticorrelation seems to suggest that the corona is both radially and vertically extended. While the QPO frequency is mostly coupled to the radial extent, the reverberation lag is coupled to both the radial and vertical extents. The idea that the corona has both radial and vertical extents is very natural, and is previously suggested by reverberation features in AGN, which require both a radially extended corona over the disk surface and a vertically extended one sitting at the launching point of a jet \citep{wilkins2016towards,wilkins2017revealing}. If the QPO is produced by the Lense-Thirring precession of an inner hot flow and/or the jet \citep{ma2021discovery,liska2021disc}, which shrinks in the radial dimension before the source reaches the soft state, the longer reverberation lags require an additional component of the corona on the vertical axis of the BH that expands during the state transition. In the narrow line Seyfert 1 galaxy Mrk 335, the emissivity profile of the reflection spectrum indicates the radially extended corona becomes gathered up in the center and is then launched away from the BH, reminiscent of a jet, during a flaring event \citep{wilkins2015driving}. This scenario is very similar to the physical picture we propose here to explain the evolution in the QPO frequency and reverberation lag in the hard state and the state transition of BHLMXBs. More detailed spectral-timing modelings assuming extended coronal geometries and incorporating QPO mechanisms are required to understand the QPO/reverberation connection.

\subsection{Outliers with long reverberation lags}\label{discussion:outliers}
%In the hard state, while the result of the soft lag amplitude being short ($\sim 1$~ms) with a decreasing trend is clear in the three {\color{black}best-sampled} sources (\maxij1820, \gx339, and MAXI~J1348--630), there are two outliers that show relatively long lags of $\sim 10$~ms (see Fig.~\ref{fig:unbinned_hue_1}). 
{\color{black}In the hard state, there are two outliers that show long soft lags of $\sim 10$~ms compared to the best-sampled sources (\maxij1820, \gx339, and MAXI~J1348--630) with $\sim 1$~ms soft lags (see Fig.~\ref{fig:unbinned_hue_1}).} 
At the same time, the upper bound of the soft lag frequency is only $\sim10$~Hz, which is 10 times smaller than that of \maxij1820\ (see Fig.~\ref{fig:unbinned_hue_2}a). These are EXO~1846--031 (Fig.~\ref{fig:exo1846}) and MAXI~J1803--298 (Fig.~\ref{fig:maxij1803}). The reason for the much longer reverberation lag amplitude in EXO~1846--031 is likely that it has the largest column density among the 10 sources in our sample ($N_{\rm H}\sim10^{23}$~cm$^{-2}$). The absorption diminishes significantly the photons at soft X-rays ($<1$~keV), and therefore the lag spectral quality is not adequate to confidently detect the reverberation lag and measure its amplitude, especially when we search for reverberation in the soft band of 0.5--1~keV. The limit of lag spectral quality can be seen from the large uncertainties in the lag-frequency spectrum where soft lag is detected, and the lag-energy spectrum below 1~keV in Fig.~\ref{fig:exo1846}. The soft lag amplitude is measured to be $16\pm11$~ms, also with a large uncertainty. In fact, we find the 0.5--1~keV band is dominated by the shelf of the response matrix with such high absorption. This means that the photons registered in these energy channels are actually the photons redistributed downward from higher energies, most likely $\sim2$--3~keV which is the peak area band with remaining signal. However, we note this effect does not make our soft reverberation lag detection artificial because an intrinsic hard lag {\color{black}(i.e., the hard band when measuring the lag-frequency spectrum, 2--5~keV, lags behind the $\sim2$--3~keV band)} remains a hard lag. Encouragingly, we observe a very promising iron lag whose energy is not affected by the high galactic absorption in EXO~1846--031. If, instead, we use the tentative iron lag amplitude, this source is no longer an outlier in our sample.

The other outlier, MAXI~J1803--298, is not affected by large galactic absorption ($N_{\rm H}\sim3\times10^{21}$~cm$^{-2}$; \citealt{homan2021nicer}), and indeed in this source, the measured lag is highly significant. The longer reverberation lag is more likely to be intrinsic to the system. Candidate explanations include that MAXI~J1803--298 contains a more massive black hole or a different disk-corona geometry. Perhaps this system is unique compared to the rest of the sample because it showed a long-lived excursion to the ``steep power-law state" \citep{ubach2021nicer}, which may suggest a different coronal geometry relative to the other sources. 

\subsection{A closer look at reverberation lags in the hard state}\label{discussion:hard_state}

\begin{figure}
\centering
\includegraphics[width=1.\linewidth]{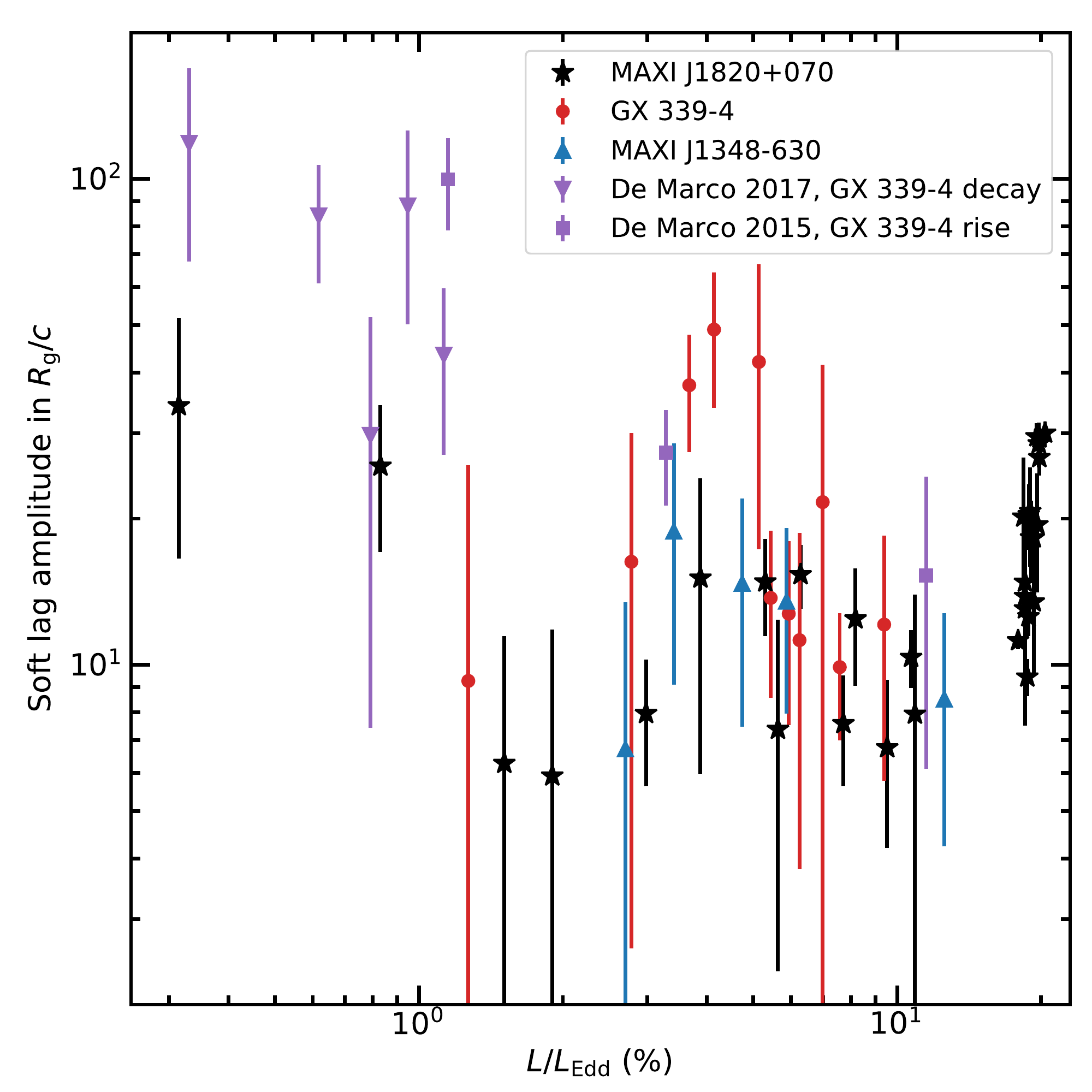}
\caption{The soft lag amplitude in light travel time (i.e., in units of $R_{\rm g}/c$) as a function of luminosity in Eddington units in the hard state. Notice the lag increase in \gx339\ could only be seen if we use the soft lags detected between 0.5--1~keV and 1.5--3~keV because the soft lag is not significant enough to detect when using 2--5~keV as the hard band. The values from previous reverberation lag measurements \citep{de2015tracing,de2017evolution} are the ``observed" lag amplitudes (without considering the dilution as shown in Table~3 in \citealp{de2017evolution}), and the $L/L_{\rm Edd}$ values are converted accordingly with our assumed BH mass and distance. See Section~\ref{discussion:hard_state} for the assumed BH mass, distance, and inclination, and how the dip results in the correlation of soft lag amplitude and hue in Fig.~\ref{fig:unbinned_hue_1} and \ref{fig:6sources_binned}(\textit{a}) when hue is $<20$ degrees. 
}
\label{fig:hard_state_lag_edd}
\end{figure}

In the hard state, we see an increase in soft lag amplitude as a function of hue when hue is $\lesssim20^\circ$ (see Figs.~\ref{fig:unbinned_hue_1} and \ref{fig:6sources_binned}a). To explore this further, we plot the soft lag amplitude in the hard state as a function of the luminosity in Eddington units\footnote{\maxij1820\ has the most reliable measurements of its BH mass ($8.48^{+0.79}_{-0.72}$~$M_\odot$; \citealp{torres2020binary}), distance (2.96$\pm$0.33~kpc; \citealp{atri2020radio}), and inclination (63$\pm$3 degrees; \citealp{bright2020extremely}). From near-infrared study, the latest measurement of the mass function of \gx339\ is $f(M)=1.92\pm0.08$~$M_\odot$, and the mass ratio of companion star and the BH is $q=0.18\pm0.05$; the distance to \gx339\ has only a lower limit of $\sim5$~kpc, and a larger distance is favored \citep{heida2017potential}. If a distance of 10~kpc and an inclination of $\sim40^\circ$ \citep{javier_gx339,wang2020relativistic} are assumed, the mass becomes 10~$M_\odot$ from the mass function. As for MAXI~J1348--630, the distance is assumed to be 3.4~kpc, the mass to be 10~$M_\odot$ \citep{lamer2021giant}, and the inclination to be $30^\circ$ (jet motion constrains $i\leq46^\circ$; \citealp{carotenuto2021black}). } for the three sources with more than one soft lag detection in the hard state (\maxij1820, \gx339, and MAXI~J1348--630) in Fig.~\ref{fig:hard_state_lag_edd}. In the initial rise of the hard state in \maxij1820's 2018 outburst, the source luminosity in \nicer's obsIDs~1200120101--1200120105 increased from 1\% to 5\%~$L_{\rm Edd}$. During this stage, the hue decreased from $-3^\circ$ to $-6^\circ$, and then increased to $2^\circ$. At the same time, the reverberation lag amplitude decreased from 25 to 6~$R_{\rm g}/c$, and then increased {\color{black}to} 15~$R_{\rm g}/c$. This correlation between hue and reverberation lag amplitude results in the increase of soft lag amplitude with hue at $\lesssim20^\circ$. After the initial rise, \maxij1820\ was in a plateau where the luminosity slightly decreased from 20\% to 18\%~$L_{\rm Edd}$, hue gradually increased from $20^\circ$ to $120^\circ$, and the soft lag amplitude decreased from 30 to 10~$R_{\rm g}/c$. After the plateau, the source went through a decay and then a rise, before finally making the hard-to-soft state transition (see e.g., Fig.~\ref{fig:hid_hue}a). During this stage, the hue spanned the range of $110^\circ$ to $140^\circ$ and correlated with luminosity (in the range of 6\% to 11\%~$L_{\rm Edd}$), while soft lag amplitude (in the range of 7 to 10~$R_{\rm g}/c$) anticorrelated with luminosity. Therefore, after the initial rise, the soft lag amplitude kept decreasing with hue in the hard state as seen at hue $\gtrsim20^\circ$ in Figs.~\ref{fig:unbinned_hue_1} and \ref{fig:6sources_binned}(a). In \gx339\ and MAXI~J1348--630, the outburst history is not as complicated, and the data groups in the hard state are all during the initial rise.

In Fig.~\ref{fig:hard_state_lag_edd}, we can see that for $L/L_{\rm Edd}\lesssim2\%$, the reverberation lag amplitude decreases with increasing luminosity, as expected from previous results. However, in the $2\%\lesssim L/L_{\rm Edd}\lesssim4\%$ range, the lag amplitude surprisingly appears to \textit{increase} with increasing luminosity, before once again decreasing with increasing luminosity for $4\%\lesssim L/L_{\rm Edd}\lesssim10\%$. Among the 3 sources, only \maxij1820\ shows the plateau stage at $\sim20\%$~$L_{\rm Edd}$, where both the luminosity and the lag amplitude steadily decrease. We also note that this finding has no conflict with previous reverberation lag measurements in \citet{de2015tracing,de2017evolution}, as the increase of lag amplitude we see in Fig. \ref{fig:hard_state_lag_edd} occurs in a luminosity range not covered in those papers.

Although the increase of reverberation lag between 2\% and 4\%~$L_{\rm Edd}$ is subtle because of the large uncertainties of the lag amplitudes, it is interesting that we see this behaviour in all 3 sources with more than 1 soft lag detection in the hard state. A possible interpretation is that below 1--2\%~$L_{\rm Edd}$, the disk is truncated at the order of several tens of $R_{\rm g}$, and above that, the disk is no longer truncated (the inner edge of the disk is only several $R_{\rm g}$), which is consistent with the spectral analysis of \gx339\ (\citealp{javier_gx339,wang2018evolution}; {\color{black}see the schematic referred to as ``faint hard state" in Fig.~\ref{fig:schematics})}. The levels of disk truncation also match the soft lag amplitudes in light travel time (in units of $R_{\rm g}/c$, see Fig.~\ref{fig:hard_state_lag_edd} below 2\%~$L_{\rm Edd}$). The change from a truncated disk to a non-truncated one is very likely to be due to the radiative efficiency of the inner disk increasing above $\sim1\%$~$L_{\rm Edd}$, which is theoretically predicted \citep{yuan2014hot} and supported by the change of slope in the radio/X-ray correlation (e.g., \citealp{koljonen2019radio}). Instead, above 2\%~$L_{\rm Edd}$, the increase and then decrease of soft lag amplitude is more likely to be driven by changes in the coronal geometry, which would be consistent with an evolving base of the compact jet. Recent GRMHD simulations \citep{liska2019bardeen} show that, as the inner disk is more radiatively efficient and thus thinner, the inflowing material exerts less pressure on the jet boundary, which becomes less collimated, resulting in the base of the jet being more extended compared to lower luminosities \citep{lucchini2021correlating,cao2021evidence}. The decreasing soft lag amplitude above 4\%~$L_{\rm Edd}$ could then be caused by a contracting X-ray corona before the quenching of the compact radio jet {\color{black}(see the schematic referred to as ``bright hard state" in Fig.~\ref{fig:schematics})}. This is similar to the expanding X-ray corona that happens before the transient radio jet ejection in the state transition. However, the interpretation should be taken with some caveats. First, the evolution of soft lag amplitudes is tentative because of the large uncertainties of the measured lags. Second, the uncertainties in BH mass, distance, and the bolometric correction factor preserve the trend, but shift the values of luminosities in Eddington units. Further detailed reflection modeling of the data in the hard state (especially measurements of the coronal height, reflection fraction, and the inner edge of the disk), and comparison with infrared/radio data are needed.

\section{Summary} \label{summary}

We have constructed a \nicer\ reverberation machine aiming to systematically search for reverberation signatures in the BH/BHC LMXBs in the \nicer\ archive, and to study the reverberation lag evolution throughout the outbursts. Our major findings are as follows: 

\begin{enumerate}
\item{Out of 26 BH/BHC LMXBs \nicer\ has observed, we find data groups with high enough signal to noise to enable a soft lag detection in 10 sources, and soft reverberation lags are detected in all 10 of these sources. Detections have previously only been reported for \maxij1820\ and \gx339, and therefore for the other 8 we report the first ever reverberation detections; increasing the sample size of BHLMXBs with reverberation lag detections from 2 to 10 (3 to 11 when accounting for detections with \xmm\ in H1743--322).}
\item{We adopt the power-spectral hue to quantify what stages the sources are in during their outbursts, and find a consistent reverberation lag evolution in the hard state and state transition for the 10 sources.}
\item{In all the 9 sources with more than 1 reverberation detection, the reverberation lag is longer and dominates at lower Fourier frequencies during the hard-to-soft transition. This reverberation lag evolution has been reported only for \maxij1820, and now with our systematic searches, our result {\color{black}suggests} this trend is a global property of state transitions of BHLMXBs that does not depend on inclination, BH mass or spin. This reverberation evolution suggests a growing X-ray emitting region, possibly due to an expanding corona.}
\item{The frequency range where we see reverberation correlates with the Type-C QPO frequency in the hard state, but they become anticorrelated during the state transition. This result hints at a corona that is both radially and vertically extended, and while the QPO frequency is mostly coupled to the radial extent, the reverberation lag is coupled to both extents.}
\end{enumerate}

%\bigskip
JW, GM, EK and JAG acknowledge support from NASA~ADAP grant 80NSSC17K0515. JAG thanks support from the Alexander von Humboldt Foundation. AI and DA acknowledge support from the Royal Society. RAR acknowledges support from NICER grant 80NSSC19K1287.

\appendix
\section{Reverberation lag detections in individual sources}

\setcounter{figure}{0}
\renewcommand{\thefigure}{A\arabic{figure}}

\setcounter{table}{0}
\renewcommand{\thetable}{A\arabic{table}}

\begin{figure}[htb!]
\centering
\includegraphics[width=0.8\linewidth]{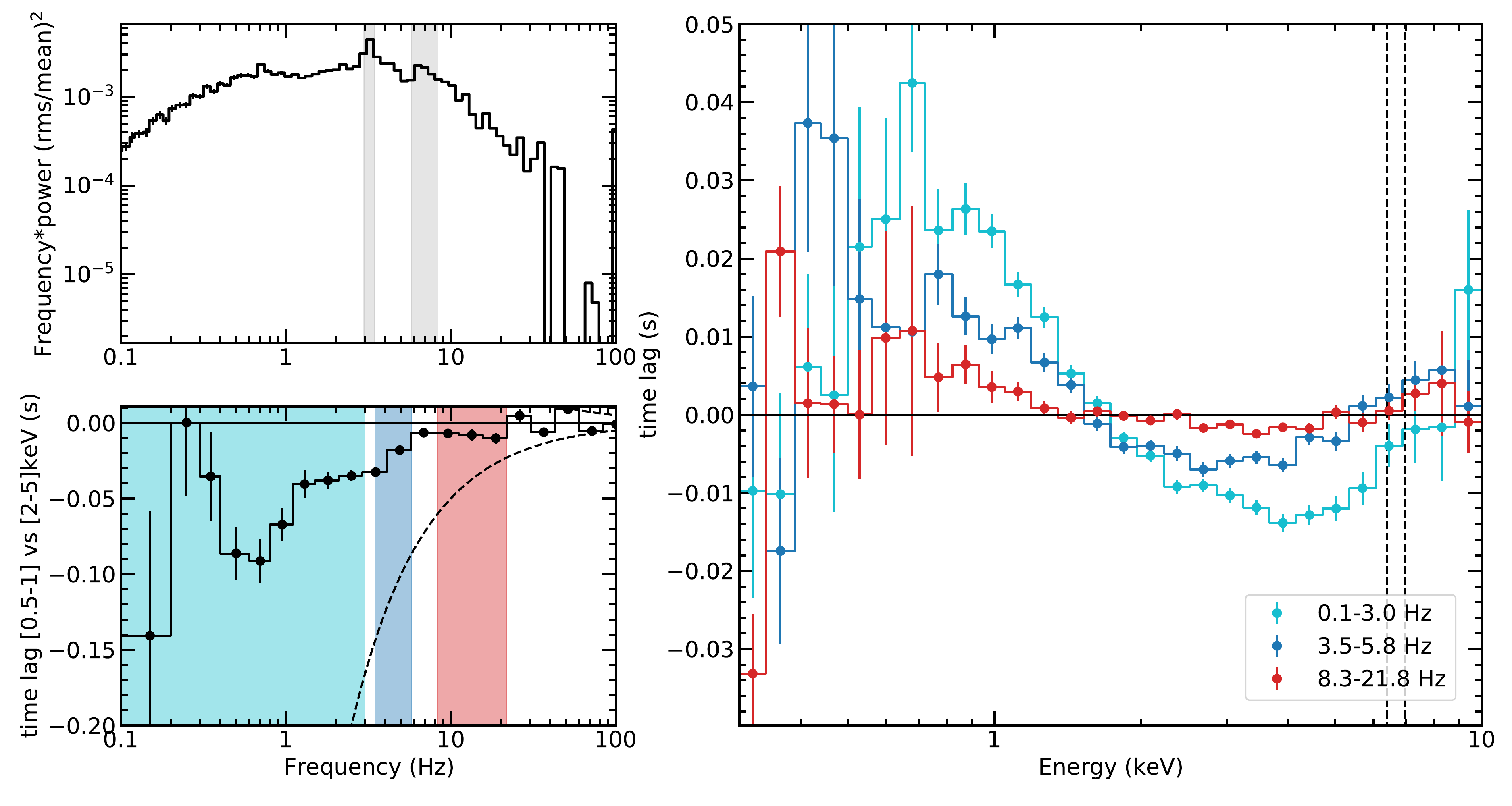}
%\caption{For the data group in the HIMS of \gx339\ (corresponding to the red group shown in Fig.~\ref{fig:gx339}), (\textit{right}) the lag-energy spectra in the unconnected frequency bands outside of the frequencies of the QPO and (sub)harmonics (central frequency$\pm$HWHM) which are shown with shaded regions in the PSD (\textit{upper left}) and the lag-frequency spectrum (\textit{lower left}). {\color{black}The unconnected frequency bands outside of the frequencies of the QPO and (sub)harmonics are 0.1--3.0~Hz, 3.5--5.8~Hz, and 8.3--21.8~Hz, which are also shown with red shaded regions in the lag-frequency spectrum.}}
\caption{{\color{black}(\textit{Right}) The lag-energy spectra of \gx339\ in the HIMS (data corresponding to the red group in Fig.~\ref{fig:gx339}) in three frequency bands chosen to exclude the QPO frequency bands. (\textit{Lower left}) The lag-frequency spectrum with the three bands (0.1--3.0~Hz, 3.5--5.8~Hz, and 8.3--21.8~Hz) shaded in colors corresponding to those of the lag-energy spectra. (\textit{Upper left}) PSD with the QPO bands (central frequency $\pm$ HWHM) shaded in grey. }}
\label{fig:gx339_qpo_treatment}
\end{figure}

There are 26 BH/BHC LMXBs observed by \nicer. Among those, the reverberation machine finds good data groups and detects soft reverberation lags in 10 sources (see Table~\ref{tab:sample_gg}). The reverberation machine summary plots include the spectral-timing results for a group in the bright hard state and a group in the intermediate state (if available). {\color{black}We note that an example of the lag-energy spectra of \gx339\ in the discontinuous frequency bands outside of QPO frequencies is shown in Fig.~\ref{fig:gx339_qpo_treatment}. In the 10 reverberation machine summary plots (Figs.~\ref{fig:gx339}, \ref{fig:maxij1820}--\ref{fig:maxij1803}), we show only the lag-energy spectra in the full Fourier frequency band in which we detect soft lags for clarity.} The results for \gx339\ are detailed in Section~\ref{results:detections}, and those for the other 9 sources are presented below. The reverberation machine finds no good data group in the other 16 sources, and the peak count rate, peak rms, total exposure, the states achieved, and the reasons for no good groups are shown in Table~\ref{tab:sample_no_gg}.

\subsection{MAXI J1820+070}
\label{maxij1820}
\maxij1820\ is one of the brightest BHLMXBs detected in history as it reached 5.8~Crab at the peak luminosity in the 2018 outburst. The brightness and great exhibition of multi-wavelength activity across spectral states make it an extensively studied new BHLMXB (e.g., \citealp{homan2020rapid,tetarenko2021measuring}). The BH mass is measured to be $8.48^{+0.79}_{-0.72}$~$M_\odot$ \citep{torres2020binary}, and the distance to be 2.96$\pm$0.33~kpc \citep{atri2020radio}. The inclination constrained from the jet motion is 63$\pm$3 degrees  \citep{bright2020extremely}.

Thanks to the high brightness and near-daily cadence coverage of this source, we find 88 good groups (making use of data with obsID~1200120101 in the initial rise of hard state to obsID~2200120309 in the quiescence), and detect soft reverberation lags in 84 of them. The mean lag significance indicator is 3115, which is the highest among all 10 BHLMXBs with detected reverberation lags. The non-detection groups lie in the lower branch of HIMS (i.e., soft-to-hard transition) approaching the hard state for the significance of promising soft lag being lower than $2\sigma$.

\begin{figure*}[htb!]
\centering
\includegraphics[width=1.\linewidth]{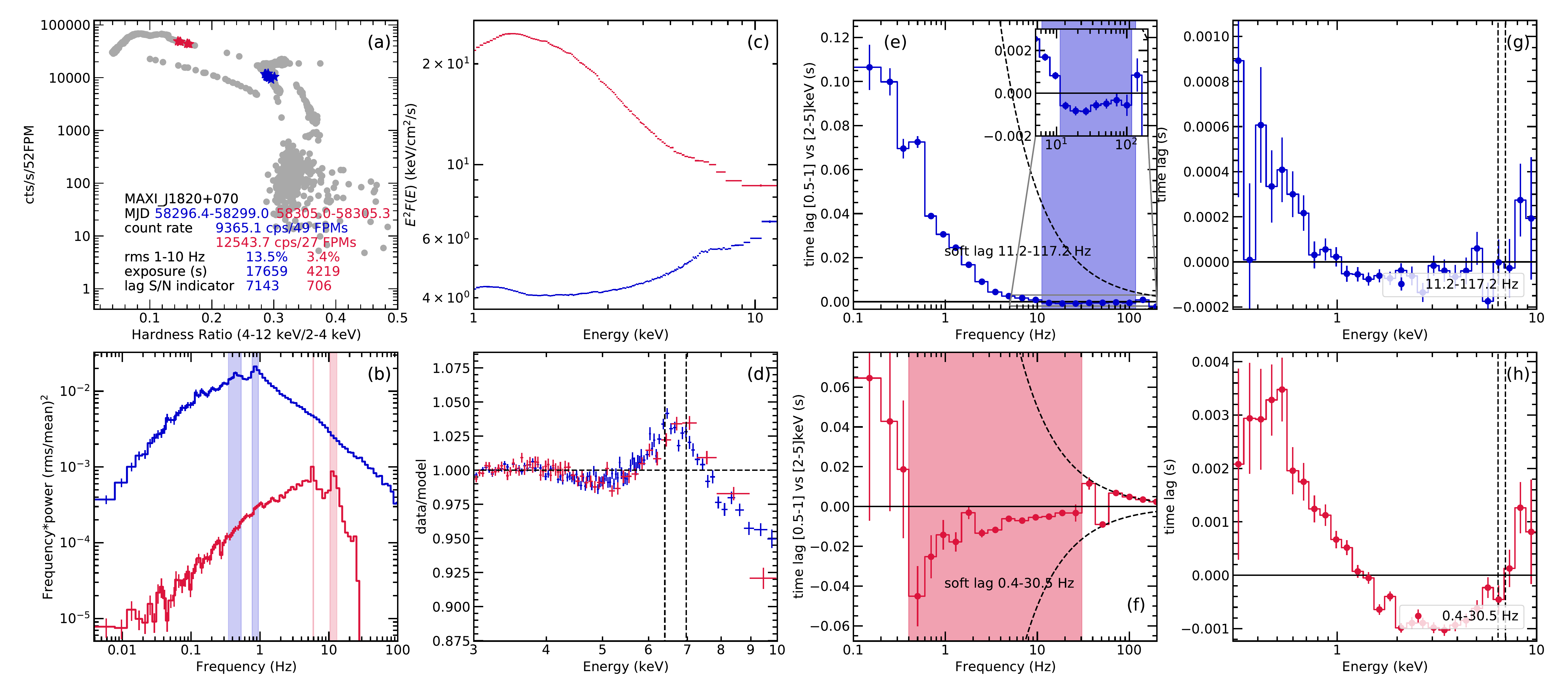}
\caption{Reverberation machine summary plot for \maxij1820\ (see Section~\ref{maxij1820} for more details). 
}
\label{fig:maxij1820}
\end{figure*}

We note that with \nicer, we have performed detailed time lag analysis for \maxij1820\ with a conventional method of {\color{black}manually} grouping the data \citep{kara2019corona, wang2021disk}. With the new automatic grouping, we verify in Fig.~\ref{fig:maxij1820} that the major result of \citet{wang2021disk}, i.e.{\color{black},} that the reverberation lag frequency {\color{black}decreases} during transition with growing lag amplitude, is well reproduced.

\begin{figure*}
\centering
\includegraphics[width=1.\linewidth]{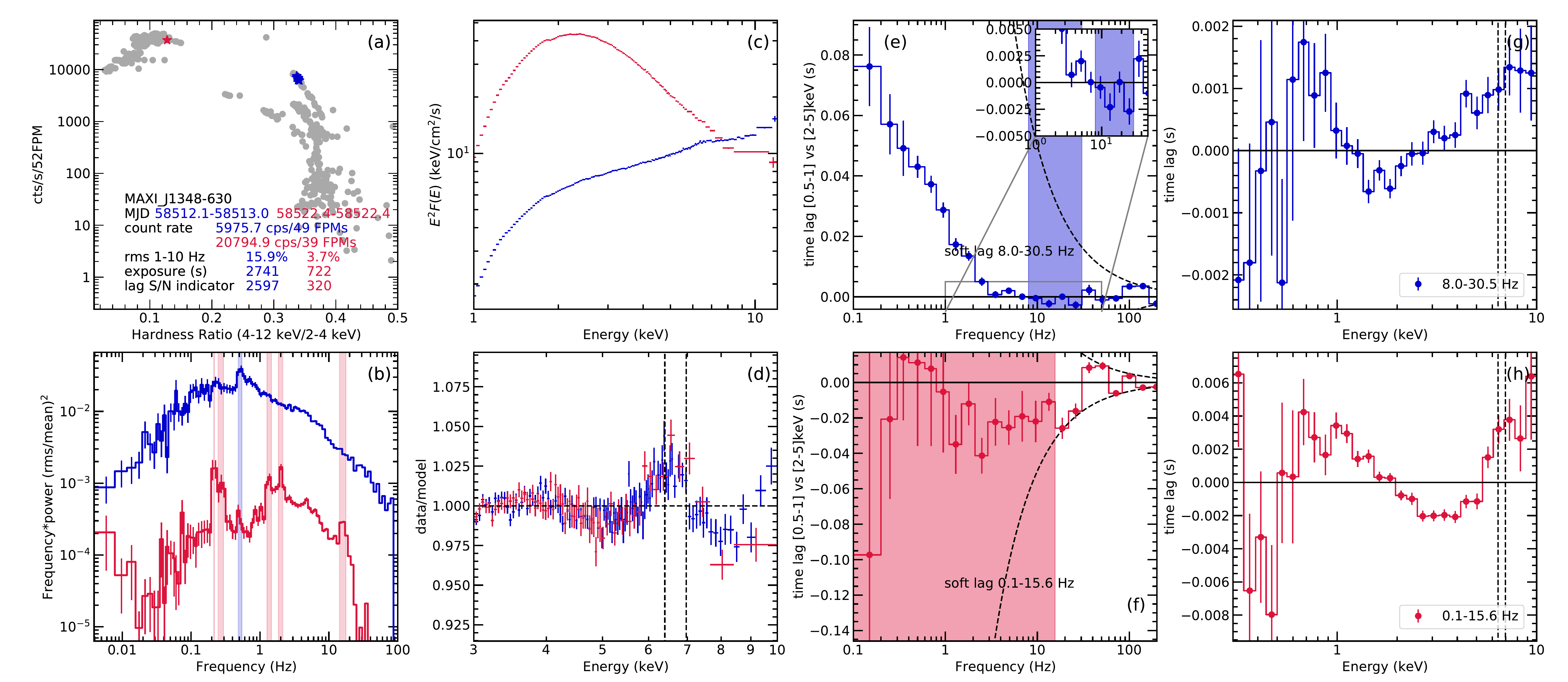}
\caption{Reverberation machine summary plot for MAXI~J1348--630 (see Section~\ref{maxij1348} for more details). 
}
\label{fig:maxij1348}
\end{figure*}

\subsection{MAXI J1348--630}
\label{maxij1348}

MAXI~J1348--630 is a bright X-ray transient reaching 4.5~Crab at its peak luminosity in its Jan 2019 outburst \citep{tominaga2020discovery}, after which 4 hard-only mini-outbursts were observed \citep{zhang2020nicer}. With H I absorption, the distance of MAXI~J1348--630 is measured to be $2.2^{+0.5}_{-0.6}$~kpc \citep{chauhan2021measuring}, and the jet motion constrains the ejecta inclination to our line of sight to be $i\leq46^\circ$ \citep{carotenuto2021black}. Moreover, MAXI~J1348--630 was believed to host a relatively massive black hole \citep{tominaga2020discovery}, but with the more recent distance measurement, the evidence has become weaker because the lower limit of $M_{\rm BH}$ is only 3.7~$M_\odot$ assuming $a_*=0$ and $i=0^\circ$ \citep{chauhan2021measuring}.

%We find 34 good groups and detect soft reverberation lags in 15 groups, among which 14 groups are during the major outburst in both hard and intermediate states, and 1 group is in the first mini-outburst. 
{\color{black}We find 34 good groups and detect soft reverberation lags in 15 groups. Among them, 14 groups are during the major outburst covering both the hard and intermediate states, and the remaining 1 group is in the first hard-only mini-outburst after the major outburst.}
The non-detections all lie in the hard state because of the intrinsic shape difference of the lag-energy spectrum compared to the HIMS. The evolution in the soft reverberation lag frequency and amplitude is consistent with what has been detected in \maxij1820 and \gx339 (see Fig.~\ref{fig:maxij1348}). We note that \nicer\ had a visibility gap for this source between MJD~58513.2 (obsID~1200530105) when it was in the bright hard state and MJD~58521.9 (obsID~1200530106) when the source was already in the HIMS, during which time a relativistic transient jet was estimated to be ejected on MJD~$58518.9\pm2.4$ \citep{carotenuto2021black}. Therefore, we cannot conclude when the state transition starts relative to the transient jet launching, nor can we precisely determine when the X-ray reverberation lag becomes longer during the state transition.

\subsection{MAXI J1535--571}
\label{maxij1535}

MAXI J1535--571 is a LMXB that was first detected by MAXI on 2017 September 2 \citep{negoro2017maxi}. By fitting the soft state spectrum with a relativistic thin disk component \texttt{kerrbb}, the black hole mass is estimated to be $10.39^{+0.61}_{-0.62}$~$M\odot$ with a distance of $5.4^{+1.8}_{-1.1}$~kpc  \citep{sridhar2019broad}. Reflection spectroscopy suggests an inner disk inclination in the range of 55--77 degrees \citep{xu2018reflection,miller2018nicer}; but the jet inclination inferred from the motion of the superluminal knot is $\leq$45 degrees \citep{russell2019disk}. Notice that MAXI~J1535--571 has a relatively high interstellar absorption with $N_{\rm H}=(3-8)\times10^{22}$~cm$^{-2}$ \citep{kennea2017maxi,gendreau2017initial,xu2018reflection}, which makes the soft lag more challenging to measure than in less obscured sources. In the radio, a transient jet knot was estimated to be ejected on MJD~$58010^{+2.65}_{-2.5}$ \citep{russell2019disk}, and the compact jet was rapidly quenching on MJD~58013.6 \citep{russell2020rapid}.

\begin{figure*}
\includegraphics[width=1.\linewidth]{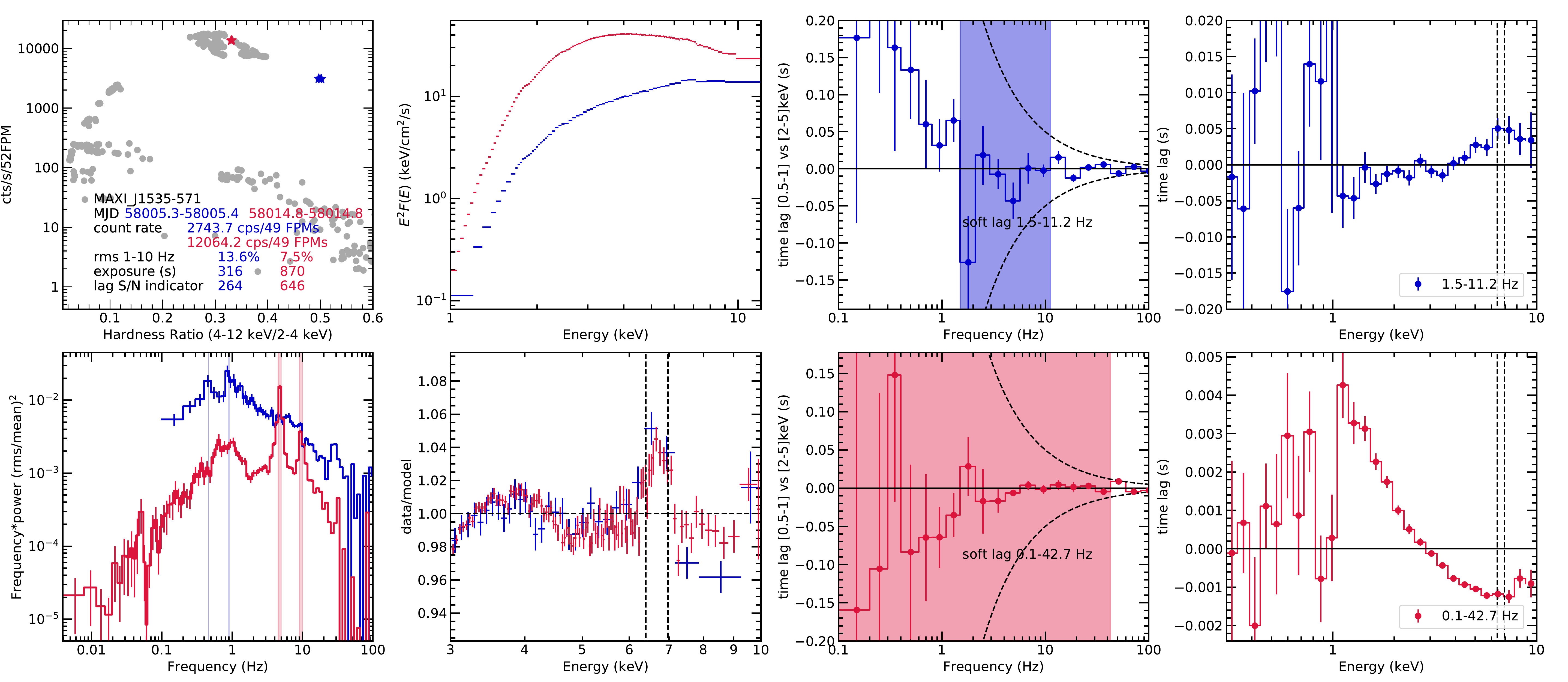}
\caption{
%MAXI J1535--571, soft lag detected in 12 out of 14 data groups; 1 group in the hard state at the beginning of \nicer\ coverage (blue), and 11 groups in the HIMS (a representative one is shown in red). 
Reverberation machine summary plot for MAXI~J1535--571 (see Section~\ref{maxij1535} for more details). The PSD of the only group in the hard state starts from 0.1~Hz because none of the GTIs in this group is longer than 256~s, and the segment length is changed to 10~s in this case.
}
\label{fig:maxij1535}
\end{figure*}

\nicer\ started observing this transient from 2017 September 9 (MJD~58005.3), corresponding to the observation (obsID~1050360103) that composes the first good group we find with our reverberation machine (see Fig.~\ref{fig:maxij1535}). 
%For this data group, though we lack a measurement of hue to determine the spectral state as no GTI in this data group is longer than 256~s to measure the power down to 0.0039~Hz, we confirm its hard state nature from spectral-timing features. 
{\color{black}In this data group, no GTI is longer than 256~s to measure the power down to 0.0039~Hz. As one ingredient in measuring the hue is the power in 0.0039--0.031~Hz, we cannot use the hue to determine the spectral state for this data group. Instead, we confirm its hard state nature from spectral-timing features.}
The QPO centroid frequency is relatively low at 0.9~Hz. The photon index is $\sim1.8$, and the disk is cool at $T_{\rm in}\sim0.18$~keV when the flux-energy spectrum is fitted in 0.5--10~keV range with a basic reflection model of \texttt{(tbabs*diskbb+reltransDCp)} following \citet{wang2021disk}. The other 13 good groups are all in the HIMS covering MJD~58008.5--58038.4 (obsID~1050360104 to 1130360115)\footnote{It was suggested that the source entered the SIMS between MJD~58014.2 and 58015.4 \citep{tao2018swift}. Meanwhile, all the good groups we find in the intermediate state are determined to be in the HIMS based on the hue. There is no controversy because by comparing with the observations classified to be SIMS with the \nicer\ data, our good groups do not cover any of those data segments (see Fig.~1 in \citealp{stevens2018nicer}). This is owing to the criteria of the relative change in fractional variance when grouping the data, and the requirement on the lag significance indicator that is most sensitive to the rms.}, and the 2 non-detections happen along with the two lowest lag significance indicators. A representative group in the HIMS is also shown in Fig.~\ref{fig:maxij1535}, and we observe again that the frequency over which soft reverberation lag dominates decreases during state transition. Because of the gap between MJD~58005.3 and 58008.5, and the large uncertainty in the transient jet ejection time (MJD~$58010^{+2.65}_{-2.5}$), we cannot determine the time between the X-ray corona expansion and the launch of the transient jet.

\subsection{MAXI J1631--479}
\label{maxij1631}

MAXI~J1631--479 was discovered as a bright hard X-ray source on 2018 December 21 by MAXI/GSC \citep{kobayashi2018maxi}, and was in the hard state until December 26 (MJD~58478; \citealp{fiocchi2020evolution}). MAXI~J1631--479 is likely to be a low-inclination source based upon several arguments including from reflection spectroscopy with \nustar\ ($i=29\pm1^\circ$; \citealp{xu2020studying}), the shape of the HID \citep{monageng2021radio}, and the phase lag of its Type-C QPO \citep{eijnden2019nicer}. Both the black hole mass and distance are currently unknown. A radio ﬂare was observed to peak on MJD~58560 to 58582 (though notably, the {\color{black}HIMS-to-soft-state} transition happened earlier, between MJD~58545 and 58560). The delay of the radio flare hints at the possibility of multiple unresolved ﬂares and/or jet–ISM interactions \citep{monageng2021radio}.

\begin{figure*}[htb!]
\hspace*{2cm}
\includegraphics[width=1.\linewidth]{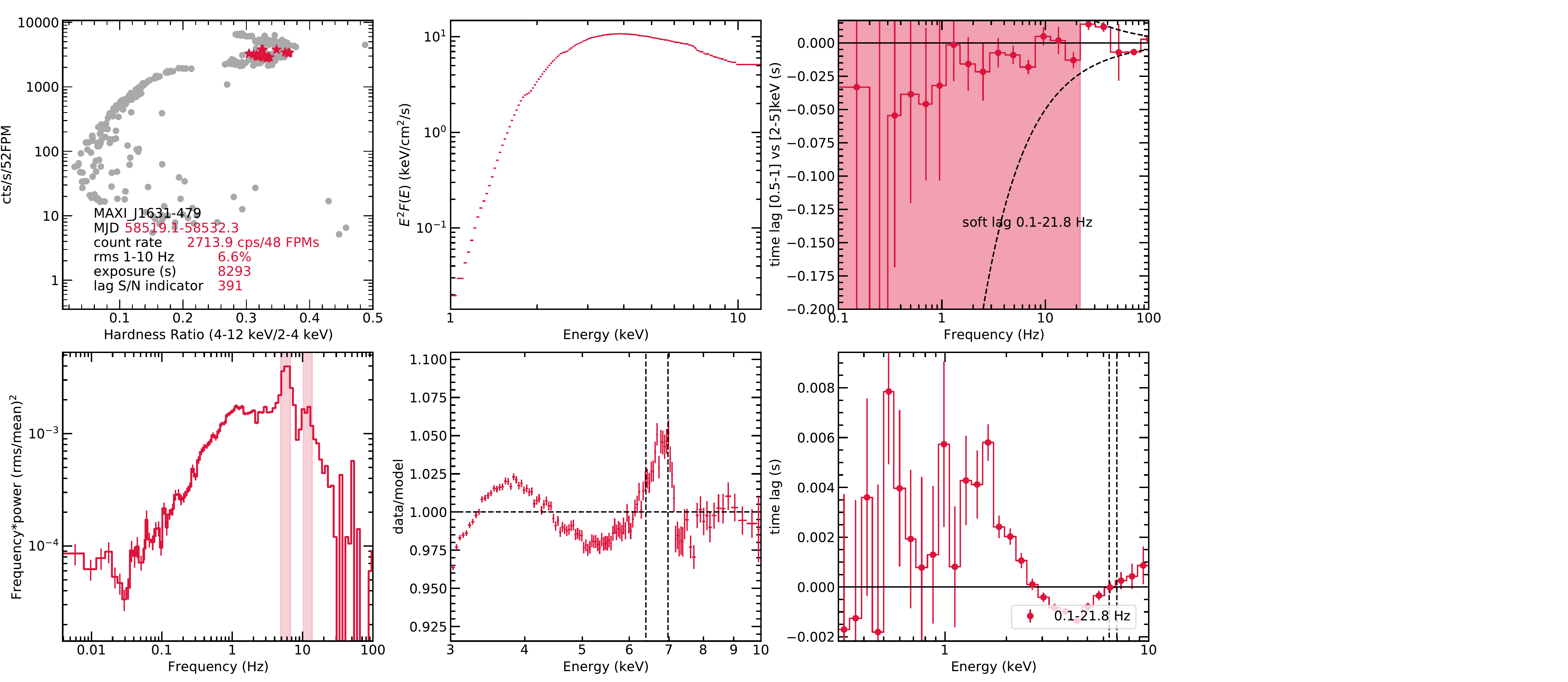}
\caption{Reverberation machine summary plot for MAXI~J1631--479 (see Section~\ref{maxij1631} for more details). All the good data groups are in the intermediate state. 
}
\label{fig:maxij1631}
\end{figure*}

\nicer\ started monitoring this source on MJD~58498 (obsID~1200500101) when it was already in the soft state, and caught a transition back to the HIMS on MJD~58506 (obsID~1200500110; \citealp{eijnden2019nicer}). The source stayed in the HIMS until MJD~58545 (obsID~2200500102), after which \nicer\ had a visibility gap and the source was already back to the soft state when the monitoring resumed on MJD~58560 (obsID~2200500108; \citealp{rout2021spectral}). As no \nicer\ data is available in the hard state, all the good groups we find are in the HIMS, and the results for one representative group is shown in Fig.~\ref{fig:maxij1631}. Soft reverberation lags are detected in 9 out of 12 good groups. These lags are always found at low Fourier frequencies (i.e., even at 0.1~Hz, no prominent hard lag is observed), consistent with results from previous sources in the HIMS.

\subsection{MAXI J1727--203}
\label{maxij1727}

MAXI~J1727--203 was discovered by MAXI/GSC on 2018 June 5 \citep{yoneyama2018maxi}. \nicer\ observed this source since the day it was discovered, and caught the initial HIMS followed by the soft state, the soft-to-hard transition, and the final hard state (see Fig.~1 in \citealp{alabarta2020x}). Spectral-timing analysis with \nicer\ data has been reported in \citet{alabarta2020x}, and the nature of the compact object in this system is by far the most uncertain.

\begin{figure*}[htb!]
\centering
\includegraphics[width=1.\linewidth]{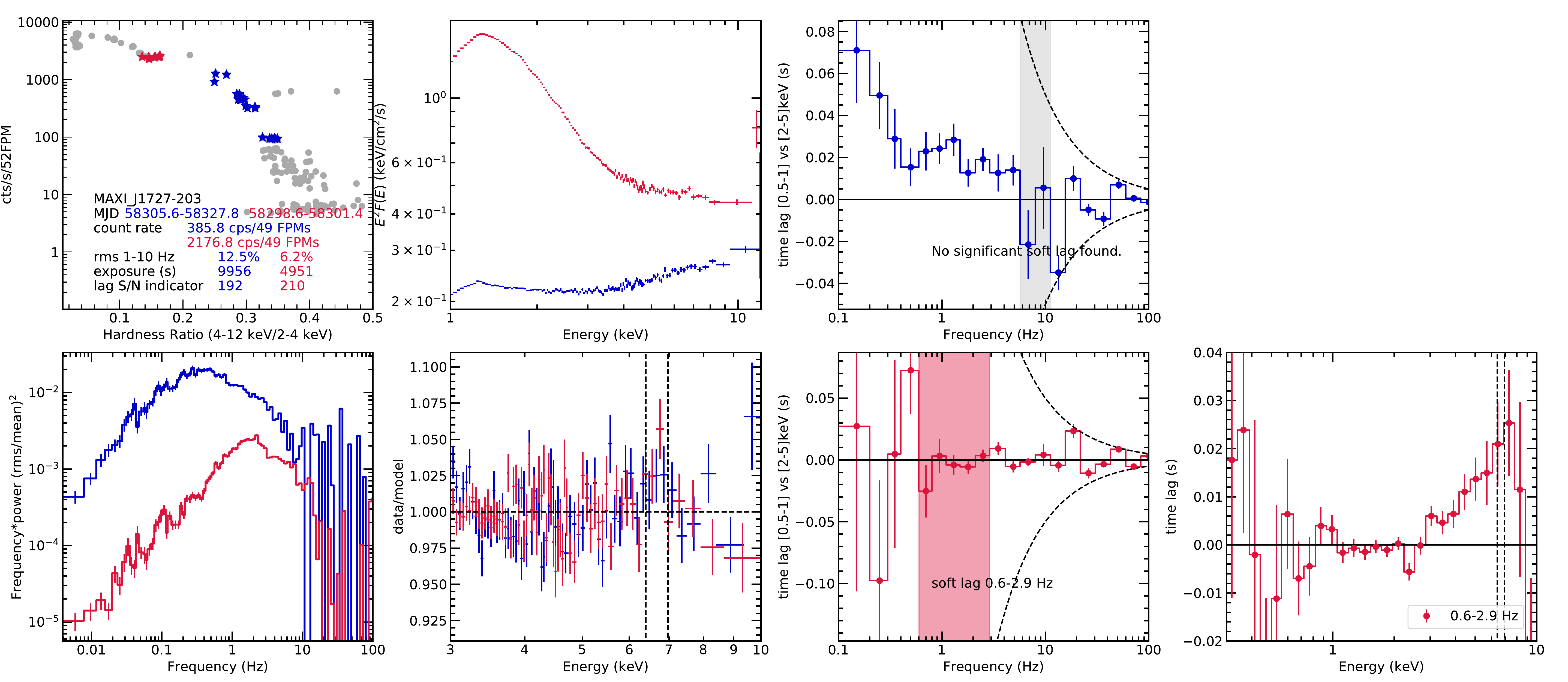}
\caption{Reverberation machine summary plot for MAXI~J1727--203 (see Section~\ref{maxij1727} for more details). 
}
\label{fig:maxij1727}
\end{figure*}

For the first time, we show here the lag analysis for this source. There are only 3 good groups found with our reverberation machine: one in the initial HIMS (MJD~58275.0--58276.5, obsID~1200220102 to 1200220103), one in the HIMS during the transition back to the hard state (MJD~58298.6--58301.4, obsID~1200220120--1200220123), and one in the hard state at the end of the outburst (MJD~58305.6--58327.8, obsID~1200220127--1200220140). The group in the hard state has only prominent low-frequency hard lags, and has no strong evidence for a high frequency soft lag, probably for the same reason as discussed in Section~\ref{results:detections} (see the blue data group in Fig.~\ref{fig:maxij1727}). The soft reverberation lag is detected in both HIMS groups (e.g., Fig.~\ref{fig:maxij1727}). {\color{black}Although the soft lag $<1$~keV is not as significant as in previous sources, we note that the iron lag is very significant in the group shown. This confirms that reverberation is present in this source. Interestingly, the relativistic iron line in the flux-energy spectrum is not as prominent as the iron lag in the lag-energy spectrum. This type of discrepancy was reported previously for several AGN, and was thought to suggest the inadequacy of the lamppost corona assumption \citep{zoghbi2020testing,mastroserio2020multi}.} The detection of its reverberation lag (especially the iron lag)
%is quite similar to other known black hole transients, 
{\color{black}hints at} a black hole nature for MAXI~J1727--203.

\subsection{EXO 1846--031}
\label{exo1846}

EXO~1846--031 was first discovered in 1985 \citep{parmar1985exo}. After 25 years in quiescence state, it underwent an outburst in 2019, and was detected first by MAXI \citep{negoro2019maxi_exo}. \citet{draghis2020new} performed reflection spectroscopy analysis with \nustar\ data, and found that EXO~1846--031 has a very high spin ($a_*\sim0.997$), a high inclination ($i\sim73^\circ$), and a high column density of $N_{\rm H}\sim(10-12)\times10^{22}$~cm$^{-2}$, which makes it difficult to dynamically measure the black hole mass in the optical. From the model \texttt{kerrbb}, the authors estimate a black hole mass of $9\pm5$~$M_\odot$. Both the extremely high spin and its position in the radio/X-ray plane \citep{miller2019vla} hint at a black hole nature for this system. 

\begin{figure*}[htb!]
\centering
\includegraphics[width=1.\linewidth]{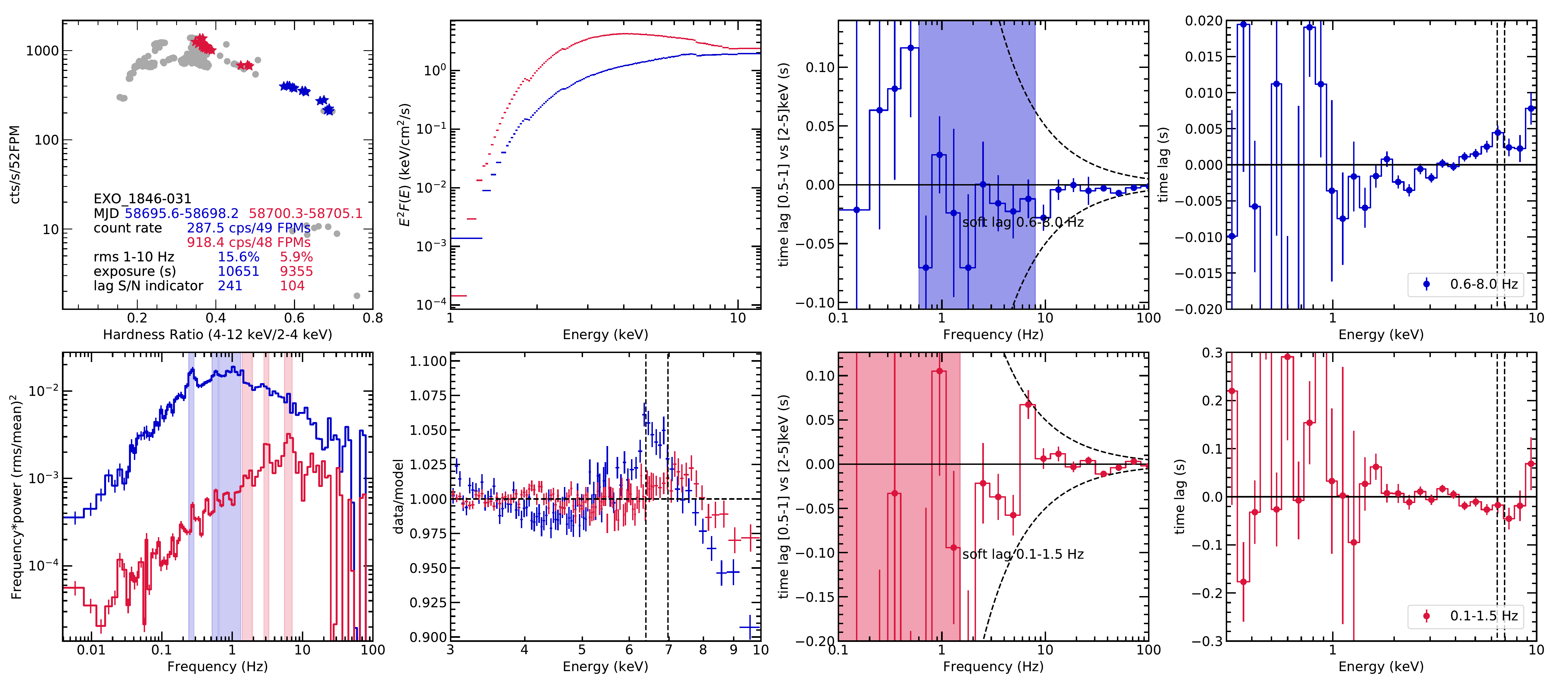}
\caption{Reverberation machine summary plot for EXO 1846--031 (see Section~\ref{exo1846} for more details). 
}
\label{fig:exo1846}
\end{figure*}

After its renewed activity, \nicer\ quickly followed up on 2019 July 31 (MJD~58695, obsID~2200760101) and confirmed the source was in the hard state \citep{bult2019nicer}. With our reverberation machine, we have found 3 good groups, 1 in the initial hard state, and 2 in the intermediate state. A soft reverberation lag is detected in the first two groups (see Fig.~\ref{fig:exo1846}). {\color{black}We note that although the soft X-ray data quality is largely affected by the large column density towards this source (see also Section~\ref{discussion:outliers}), an iron lag is present in the lag-energy spectrum in the hard state data group. On the other hand, the shape of the lag-energy spectrum for HIMS data group is certainly not a typical hard lag (which would monotonically increase with energy without any feature). Therefore, we confirm that reverberation is present in these two data groups.} By comparing results from these two groups, we see the frequency over which soft reverberation lag is detected decreases during the hard-to-soft transition, and the iron line seems to get blueshifted, exactly as seen previously in \maxij1820 \citep{wang2021disk}. The nature of the blueshifting is not well understood, and physically-driven models {\color{black}that are preliminarily promising} include the emission from the plunging region \citep{fabian2020soft} and returning radiation \citep{connors2020evidence}. {\color{black}Other possibilities include extra Comptonization of the reﬂection spectrum in the corona \citep{steiner2017self}, a change in the inclination (i.e., warps) in the reflecting surface of the inner disk, a change in the disk thickness as the mass accretion rate rises, or a change in the irradiation profile due to changing coronal geometry.}

\subsection{GRS 1915+105} \label{grs1915}

\begin{figure*}[htb!]
\hspace*{2cm}
\includegraphics[width=1.\linewidth]{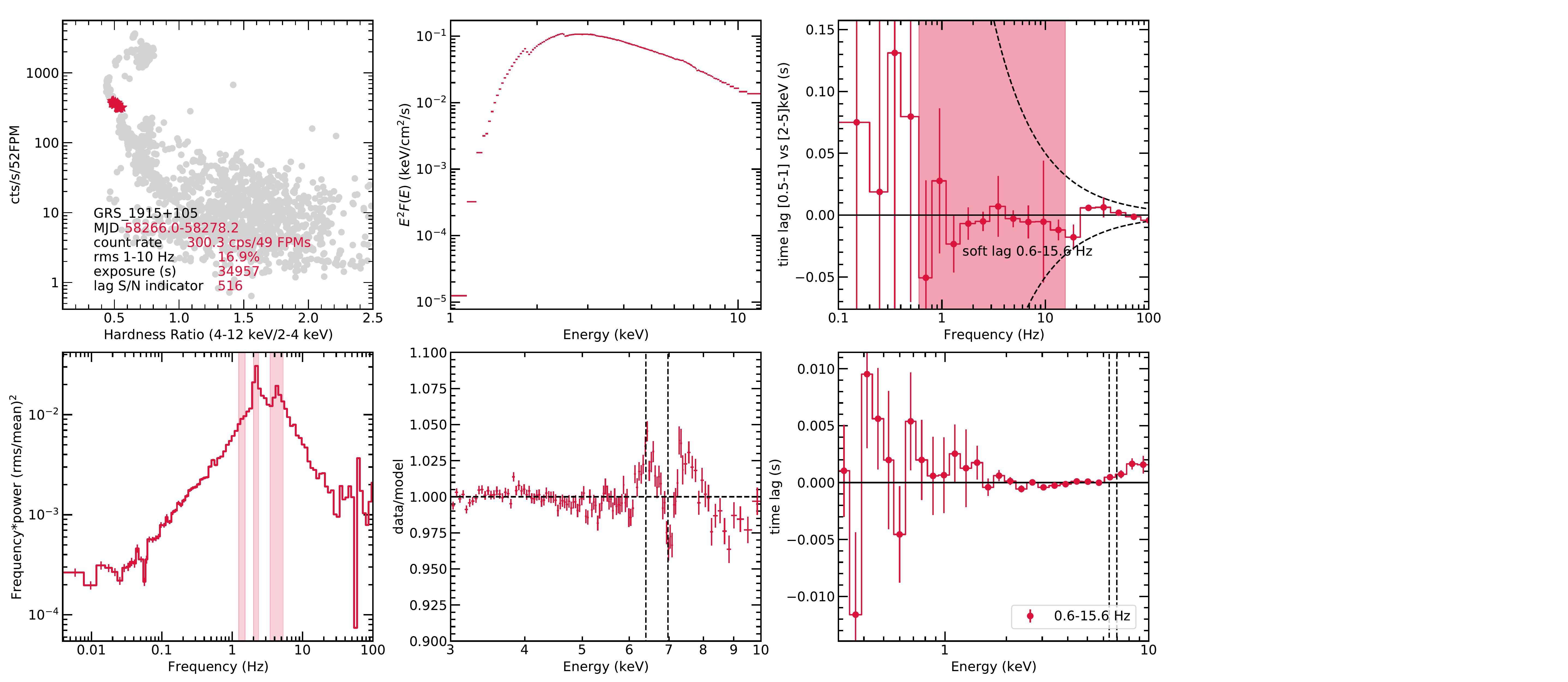}
\caption{Reverberation machine summary plot for GRS~1915+105 (see Section~\ref{grs1915} for more details). All the good data groups are in the HIMS determined from the power-spectral hue. 
}
\label{fig:grs1915}
\end{figure*}

GRS~1915+105 is the first source of detected superluminal jets in our galaxy, and therefore is the first ``microquasar" \citep{mirabel1994superluminal}. It displays at least a dozen of variability states that are believed to result from disk-jet coupling (e.g., \citealp{2000A&A...355..271B}). The BH mass is measured to be $12\pm2$~$M_\odot$, with an orbital period of 33.5~days \citep{reid2014grs1915}. GRS~1915+105 is a peculiar BHLMXB as it remained in a bright outburst for 26 years since its discovery in 1992 \citep{castro1992grs}. Since its launch, \nicer\ frequently observed GRS~1915+105 in 2017 (until 2017-11-06, obsID~1103010134). This dataset was analysed in \citet{neilsen2018persistent} and the well-known disk wind evidence including Fe~XXV and Fe~XXVI absorption was detected. From 2018 April to 2020 March, GRS~1915+105 surprisingly declined in count rate, and eventually settled in a very faint and hard state \citep{negoro2018maxi,homan2019sudden}. Very recently, in 2021 July, the source was detected to be rebrightening \citep{neilsen2021rebrightening}.

Our reverberation machine results in 3 good data groups all during the declining stage (making use of obsID~1103010147 to 2596010901, corresponding to 2018-05-27 to 2019-05-01). A soft reverberation lag is detected in the first two groups. The results of the first group is shown in Fig.~\ref{fig:grs1915}. {\color{black}The lag spectral quality is limited, and the soft lag amplitude is measured to be $8\pm5$~ms, with a relatively large uncertainty. However, we note that this measurement is within uncertainties with \maxij1820\ (see Fig.~\ref{fig:unbinned_hue_1}).} We find a broad emission line at 6.4~keV and a prominent narrow absorption feature at 6.97~keV which corresponds to Fe~XXVI line from highly ionized wind, consistent with previous \nicer\ result in the more luminous state \citep{neilsen2018persistent}. We also note that the hue for these groups suggests that GRS~1915+105 was in the HIMS in the decay stage.

\subsection{AT2019wey} \label{at2019wey}

AT2019wey was initially discovered as an optical transient in December 2019 \citep{2019TNSTR2553....1T}, and later was classified as a galactic XRB for its strong X-ray emission and hydrogen lines detected at redshift $z = 0$ \citep{yao2020x}. The source is expected to be a LMXB because the mass of the companion star was constrained to be $<0.8$~$M_\odot$ \citep{yao2020multi}. The multi-wavelength (radio to X-ray) spectral-timing features are consistent with a BHLMXB nature for AT2019wey \citep{yao2020multi,mereminskiy2021peculiar}. A resolved radio source was detected and could be interpreted as the steady compact jet common in the BHLMXB hard state \citep{yadlapalli2021vlba}. Currently, we lack a mass estimate for this source, and the inclination is constrained to be $i\leq30^\circ$ from reflection spectroscopy \citep{yao2020comprehensive}.

\begin{figure*}
\hspace*{2cm}
\includegraphics[width=1.\linewidth]{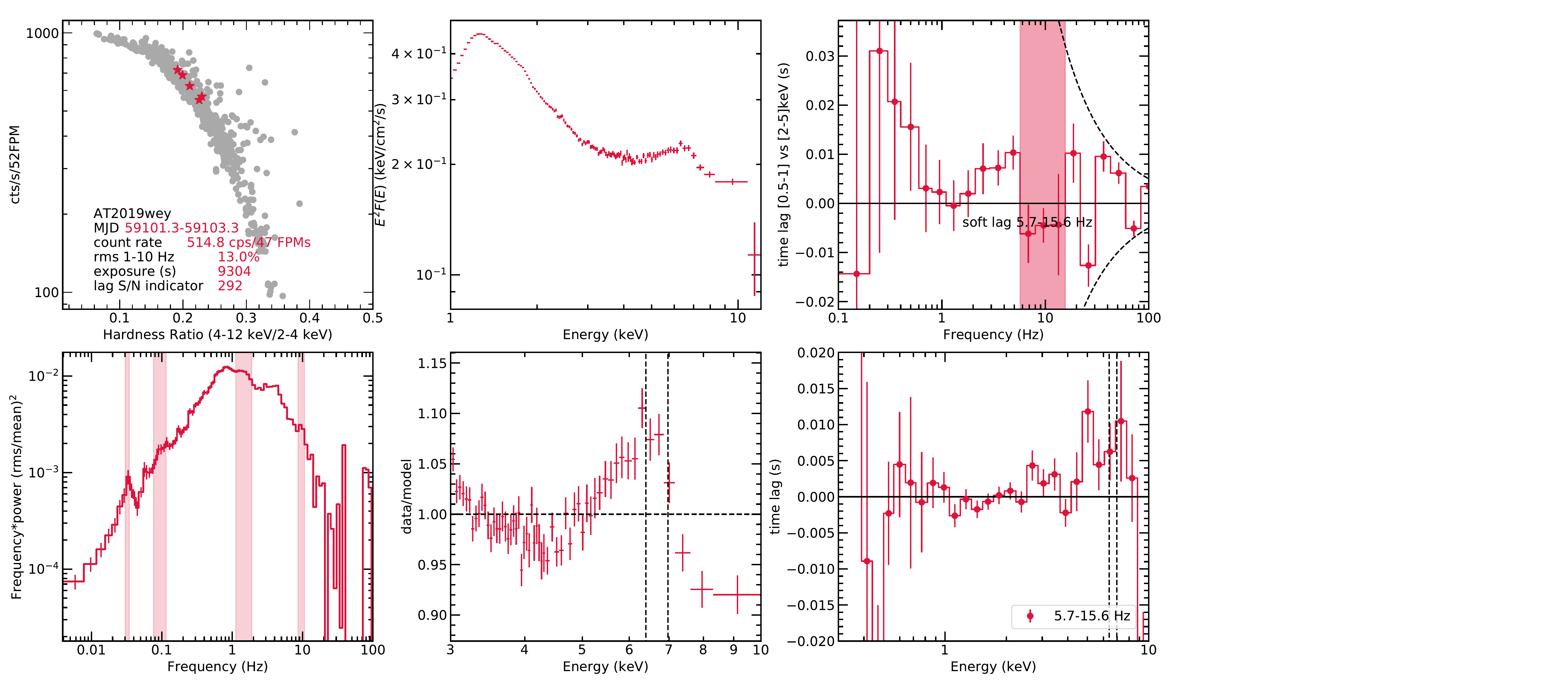}
\centering
\includegraphics[width=1.\linewidth]{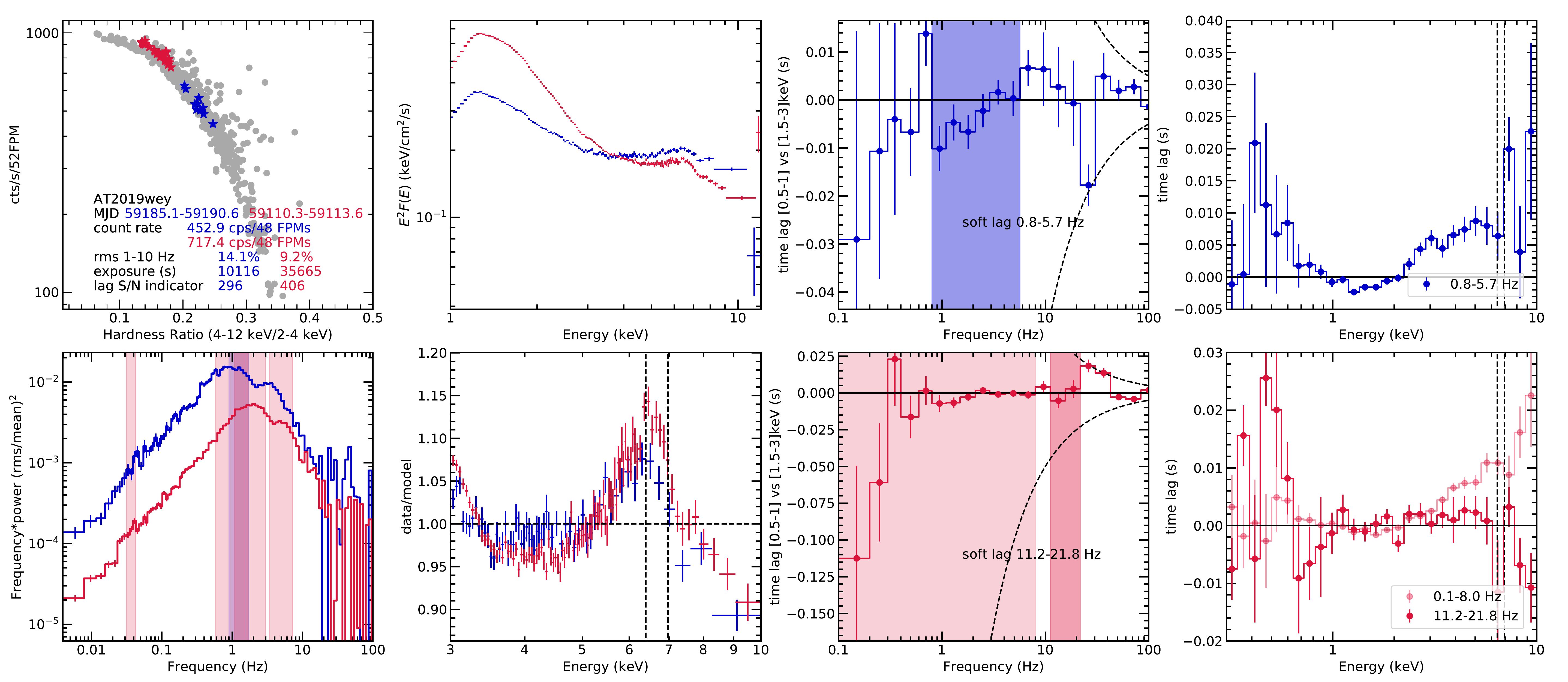}
\caption{Reverberation machine summary plot for AT2019wey. The soft lag is detected in 3 out of 22 good data groups, but the reverberation nature is difficult to confirm because of a lack of the iron line feature in the lag-energy spectrum (\textit{upper}). If we use 1.5--3~keV instead of 2--5~keV as the hard band in calculating the lag-frequency spectrum, the lag-energy spectra show more promising reverberation lag features (\textit{lower}, see Section~\ref{at2019wey} for more details). 
}
\label{fig:at2019wey}
\end{figure*}

During the \nicer\ coverage, AT2019wey was in the hard state at the beginning from MJD~59070 to MJD~59082 (obsID~3201710105--3201710117), and then transitioned into the HIMS as reported by \citet{yao2020comprehensive}, which is in great agreement with spectral states classified by the hue in our reverberation machine. Though we have found in total 22 good groups (6 in the hard state and 16 in the HIMS), the fraction of good groups with soft reverberation lag detected is by far the lowest: only 3 out of 22, all in the HIMS. In addition, the lag-energy spectra found in the soft lag frequencies this way are not representative of a soft reverberation lag (i.e., there is no strong evidence for reverberation in either the soft X-ray band or the iron line energies, see the upper panel in Fig.~\ref{fig:at2019wey}). If we use 1.5--3~keV instead of 2--5~keV as the hard band in calculating the lag-frequency spectrum, the lag-energy spectra show quite promising reverberation lag features (see the lower panel in Fig.~\ref{fig:at2019wey}). The detection of reverberation lag may suggest that this source is a BHLMXB. 

The reason for a very low reverberation lag detection fraction and a better-suited hard band of 1.5--3~keV instead of 2--5~keV is again intrinsically the shape of the lag-energy spectrum. This would mean that even in the HIMS where we observe a prominent reverberation lag in the other sources when we match the hard band of lag-frequency measurement to the energy of the ``valley" in the lag-energy spectrum (2--5~keV), AT2019wey still has a lag-energy spectrum shape similar to the other sources in the hard state (valley at 1.5--3~keV). This would imply a lower disk temperature, a lower ionization, or a lower disk density. We note that this system has a very short orbital period ($<8.6$~hours), and a low mass accretion rate that could result in the ``hard-only" outburst implied in \citet{yao2020comprehensive}.

\subsection{MAXI J1803--298}
\label{maxij1803}
MAXI~J1803--298 was first discovered by MAXI/GSC in May 2021 and quickly followed up by \nicer\ \citep{serino2021maxi,gendreau2021nicer}, making it the most newly discovered XRB in our sample. It was initially detected in the hard state between MJD~59336 and 59340 (obsID~4202130101--4202130104), and was already in the intermediate state on MJD~59352.7 (obsID~4202130105) when \nicer\ came out of a visibility gap  \citep{homan2021nicer,ubach2021nicer}. With the promising nearly-periodic X-ray absorption dips seen by \nicer\ and \nustar, and the observed disk wind, it is likely to be a high-inclination source ($i>60^\circ$), and the orbital period is between 7 and 8 hours \citep{homan2021nicer,xu2021nustar,miller2021disk}. 

\begin{figure*}[htb!]
\centering
\includegraphics[width=1.\linewidth]{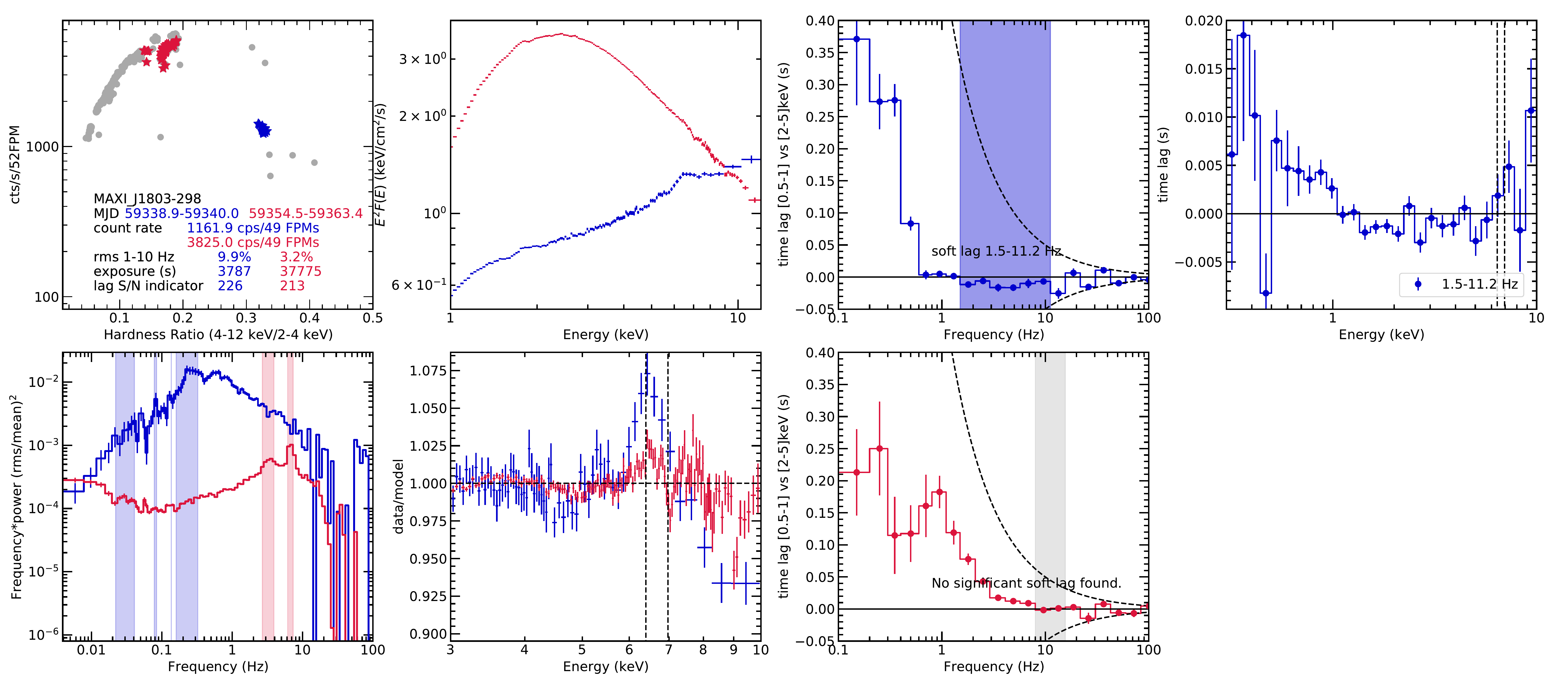}
\caption{Reverberation machine summary plot for MAXI~J1803--298 (see Section~\ref{maxij1803} for more details). 
}
\label{fig:maxij1803}
\end{figure*}

There are 2 good groups in our reverberation machine: one in the hard state (obsID~4202130103--4202130104), and the other in the HIMS (obsID~4202130107, very close to the SIMS judging by the value of hue), both are shown in Fig.~\ref{fig:maxij1803}. The reverberation lag is detected only in the hard state group, while prominent low-frequency hard lags are observed in the intermediate state group. Notice that the intermediate state group (MJD~59354.5-59363.4) has a type-B QPO detected at 6.7~Hz with a sub-harmonic at 3.3~Hz, which is consistent with what was reported in \citet{ubach2021nicer}. It was surprising at first to see that only in MAXI~J1803--298, the low-frequency hard lag remains evident in the intermediate state, where for previous sources, we see the reverberation lag dominates over the hard lags at the low frequencies. One possibility is that the intermediate state here is different from the canonical HIMS, and rather, it is in the steep power-law (SPL) state \citep{remillard2006x}, as suggested by \citet{ubach2021nicer} based on a relatively high scattering fraction. In Fig.~\ref{fig:maxij1803}, besides the blue-shifted iron line in the intermediate state, we also find evidence for absorption feature in the Fe band at 6.97~keV, suggesting the system is observed closer to edge-on {\color{black}as the disk wind is expected to be confined to the equatorial plane}, in agreement with what was seen with \swift\ \citep{miller2021disk}. 

\begin{table*}[htb!]
\begin{center}
\caption{The BH/BHC LMXB samples observed by \nicer\ without any good reverberation data from our reverberation machine, sorted by the peak count rate. \label{tab:sample_no_gg}}
\footnotesize
\begin{tabular}{cccccc}\hline \hline
Name& Peak count rate (counts/s) & Peak rms (\%) &Exposure (ks)&States achieved& Reasons for no good groups\\
\hline
4U 1543--475 & 95864 & 0.8 & 109.9 &  S & Low rms\\% the 2021 June super bright source
MAXI J0637--430 & 6452 & 36 & 155.6 &  H,I,S & Low count rate/rms\\% from Ron's Atel
4U 1957+11 & 1958 & 3 & 73.8 & S  & Low rms\\%from watchdog
Swift J1728.9--3613 & 1683 & 13  & 135.4 &  I,S & Low rms/short exposure\\%from John and Erin's review in prep
4U 1630--472 & 833 & 12  & 152.0 & H,I,S & Low rms\\%from watchdog
Swift J1842.5--1124 & 693 & 15 & 9.0 & H,I/S? & Low count rate/rms\\ % don't know the states for obs 101 and 201...have the 0.3-0.3-0.5-20 dic, gonna check the hue. %from watchdog
MAXI J1810--222 & 602 & 24  & 108.7  & H,I,S  & Low count rate/rms\\%from John and Erin's review in prep
Swift J1658.2--4242 & 342 & 19 & 62.1 &  H,I,S  & Low count rate/rms\\%from John and Erin's review in prep
XTE J1908+094 & 210  & 28 & 14.0  &  H,S  & Low count rate/rms\\%from watchdog
MAXI J1813--095 & 169 & 19 & 5.9  & H  & Short exposure\\%from John and Erin's review in prep
H1743--322 & 140 & 24 & 26.6 & H & Low count rate/rms\\%from watchdog
GRS 1758--258 & 72 & 13 & 2.9 & H & Low count rate/rms\\%from watchdog
XTE J1859+226 & 23  & 49 & 37.3  &  H  & Low count rate\\%from watchdog
IGR J17379--3747 & 20 & 28 & 136.6  &  H  & Low count rate\\%from watchdog
Swift J1357.2--0933 & 7 & 28 & 148.5 & H   & Low count rate\\%from watchdog
GRS 1739--278 & 5 & 22 & 5.6 &  H  & Low count rate\\%from watchdog
%XTE J1812--182 & 111  & 32 & 32.1  & H?  & low count rate\\ NS from 2019 detection of type I X-ray burst%from watchdog
%Swift J1858.6--0814 & & &  &  & H,I,S & low $F_{\rm var}$\\ NS from Buisson 2020b%from John and Erin's review in prep
\hline
\end{tabular}
\\
\raggedright{\textbf{Notes.} \\The count rates are for $0.3-12$~keV, and are normalized for 52~FPMs. The rms is calculated also in $0.3-12$~keV, with the Fourier frequency range of 1--10~Hz. }
\end{center}
\end{table*}

\bibliographystyle{apj}
\bibliography{resubmitted_2}
\end{document}